\documentclass[11pt,a4paper]{article}
\pdfoutput=1

\usepackage{jheppub}
\usepackage{latexsym}
\usepackage{multirow}
\usepackage{color}
\usepackage[usenames,dvipsnames,table]{xcolor}
\usepackage{graphicx}
\usepackage{epsfig}  
\usepackage{epsf}    
\usepackage{dcolumn}
\usepackage{bm}
\usepackage{dcolumn}
\usepackage{textcomp}
\usepackage{float}
\usepackage{subfig}
\usepackage{hypcap}
\usepackage[]{hyperref}
\usepackage{makecell}
\usepackage{color}
\usepackage{pifont}
\usepackage{appendix}
\usepackage{amsmath}

\hypersetup{
  bookmarks=true,         
  unicode=false,          
  pdftoolbar=true,        
 pdfmenubar=true,        
 pdffitwindow=true,     
 pdfstartview={FitH},    
 pdfsubject={Neutrino Oscillations Phenomenology},   
 pdfnewwindow=true,      
 pdfcreator={RevTeX},
 colorlinks=true,       
 linkcolor=red,          
 citecolor=blue,        
 filecolor=black,      
 urlcolor=blue,           
  }

\newcommand{\be}{\begin{equation}}
\newcommand{\ee}{\end{equation}}
\newcommand{\ba}{\begin{eqnarray}}
\newcommand{\ea}{\end{eqnarray}}

\newcommand{\dm}{\Delta m^2}

\newcommand{\dcp}{\delta_{\textrm{CP}}}

\newcommand{\dmssn}{\Delta m^2_{41}}

\newcommand{\capdef}{}
\newcommand{\mycaption}[2][\capdef]{\renewcommand{\capdef}{#2}
       \caption[#1]{{\footnotesize #2}}}
\makeatletter
\renewcommand{\fnum@table}{\textbf{\tablename~\thetable}}
\renewcommand{\fnum@figure}{\textbf{\figurename~\thefigure}}
\makeatother

\preprint{IP/BBSR/2018-2, TIFR/TH/18-07}

\title{Active-sterile neutrino oscillations at INO-ICAL \\
  over a wide mass-squared range}

\author[a,b]{Tarak Thakore,}
\author[c]{Moon Moon Devi,} 
\author[d,e,f]{Sanjib Kumar Agarwalla,}
\author[g]{Amol Dighe} 

\affiliation[a]{Louisiana State University, Baton Rouge, Louisiana 70803, USA}
\affiliation[b]{Instituto de F\`isica Corpuscular, CSIC -- Universitat de
  Val\`encia, c/ Catedr\`atico Jos\`e Beltr\`an 2, E-46980 Paterna,
  Valencia, Spain}
\affiliation[c]{Department of Physics, Tezpur University, Assam 784028, India}
\affiliation[d]{Institute of Physics, Sachivalaya Marg, Sainik School Post,
  Bhubaneswar 751005, India}
\affiliation[e]{Homi Bhabha National Institute, Anushakti Nagar,
  Mumbai 400085, India}
\affiliation[f]{International Centre for Theoretical Physics,
  Strada Costiera 11, 34151 Trieste, Italy}
\affiliation[g]{Tata Institute of Fundamental Research,
  Homi Bhabha Road, Colaba, Mumbai 400005, India}

\emailAdd{tarak.thakore@ific.uv.es}
\emailAdd{devimm@tezu.ernet.in}
\emailAdd{sanjib@iopb.res.in}
\emailAdd{amol@tifr.res.in}

\abstract
{We perform a detailed analysis for the prospects of detecting active-sterile 
oscillations involving a light sterile neutrino, over a large $\Delta m^2_{41}$ 
range of $10^{-5}$ eV$^2$ to $10^2$ eV$^2$, using 10 years of atmospheric 
neutrino data expected from the proposed 50 kt magnetized ICAL detector 
at the INO. This detector can observe the atmospheric $\nu_{\mu}$ and 
$\bar\nu_{\mu}$ separately over a wide range of energies and baselines, 
making it sensitive to the magnitude and sign of $\Delta m^2_{41}$ 
over a large range. If there is no light sterile neutrino, ICAL can place 
competitive upper limit on $|U_{\mu 4}|^2 \lesssim 0.02$ at 90\% C.L. 
for $\Delta m^2_{41}$ in the range $(0.5 - 5) \times 10^{-3}$ eV$^2$. 
For the same $|\Delta m^2_{41}|$ range, ICAL would be able to determine 
its sign, exploiting the Earth's matter effect in $\mu^{-}$ and $\mu^{+}$ events
separately if there is indeed a light sterile neutrino in Nature.
This would help identify the neutrino mass ordering in the four-neutrino
mixing scenario.
}

\keywords{Atmospheric Neutrinos, Sterile Neutrinos, ICAL, INO, Muon, Hadron}
\arxivnumber{1804.09613}

\begin{document}
\maketitle
\flushbottom

\section{Introduction and Motivation}
\label{introduction}

Unraveling neutrino properties has become an ongoing enterprise in the
intensity frontier of the high energy particle physics, both experimentally
and theoretically~\cite{Mohapatra:2005wg,Strumia:2006db,GonzalezGarcia:2007ib}. 
Active attempts are being made to probe their masses, mixings, interactions, 
Dirac vs. Majorana nature, and so on~\cite{Patrignani:2016xqp}. 
Over the last two decades, several word-class experiments involving 
solar~\cite{Cleveland:1998nv,Kaether:2010ag,Abdurashitov:2009tn,Hosaka:2005um,Cravens:2008aa,Abe:2010hy,Abe:2016nxk,Aharmim:2011vm,Bellini:2011rx,Agostini:2017ixy}, 
atmospheric~\cite{Abe:2017aap,Aartsen:2017nmd}, reactor~\cite{Gando:2013nba,An:2016ses,Abe:2014bwa,Minotti:2017kon,Ahn:2012nd,Pac:2018scx}, 
and accelerator~\cite{Ahn:2006zza,Adamson:2013ue,Adamson:2013whj,Abe:2017uxa,Abe:2017bay,Adamson:2017qqn,Adamson:2017gxd} 
neutrinos have firmly established neutrino flavor oscillations,
an engrossing example 
of a quantum mechanical phenomenon working at the macroscopic scale. 

Most of the data from the above mentioned experiments fit 
well into the standard three-flavor oscillation picture of 
neutrinos~\cite{Esteban:2016qun,Capozzi:2017ipn,deSalas:2017kay}.
The three-neutrino (3$\nu$) oscillation scheme is characterized
by six fundamental
parameters: (i) three leptonic mixing angles ($\theta_{12}$, $\theta_{13}$,
$\theta_{23}$), (ii) one Dirac CP phase ($\delta_{\rm CP}$), and 
(iii) two independent mass-squared differences\footnote{
We define $\Delta m^2_{ij} \equiv m_i^2 - m_j^2$, where $m_i$'s
    are the masses of the neutrino mass eigenstates $\nu_i$'s,
    arranged in the decreasing order of electron flavour content.
    To address the solar neutrino anomaly, we need $\Delta m^2_{21}
\approx$ $7.5 \times 10^{-5}$ eV$^2$ and to resolve the atmospheric 
neutrino anomaly, we require $|\Delta m^2_{32}|
\approx$ $2.5 \times 10^{-3}$ eV$^2$.}
($\Delta m^2_{21}$ and $\Delta m^2_{32}$).
However, not all the oscillation data are compatible with this
three-flavor oscillation picture~\cite{Abazajian:2012ys}.
There are anomalous experiments
(for recent reviews see~\cite{Collin:2016aqd,Gariazzo:2017fdh}),
which point towards oscillations with substantially large 
mass-squared difference ($\Delta m^2 \sim 1$ eV$^2$)
as compared to the well-known solar and atmospheric mass splittings. 
For example, the LSND experiment observed the appearance of $\bar\nu_e$ events
in a $\bar\nu_{\mu}$ beam, which can be explained in a two-flavor
oscillation framework with 
$\Delta m^2 \sim 1$ eV$^2$~\cite{Athanassopoulos:1995iw,Aguilar:2001ty}.
Another indication came from the short-baseline (SBL)
Gallium radioactive source experiments 
GALLEX~\cite{Kaether:2010ag} and SAGE~\cite{Abdurashitov:2009tn}, 
where the disappearance of $\nu_e$ was observed. 
Several SBL reactor antineutrino experiments noticed a
deficit in the observed $\bar\nu_e$ event rates~\cite{Mention:2011rk} 
in comparison with that expected from the calculation of reactor 
antineutrino fluxes~\cite{Mueller:2011nm,Huber:2011wv}.
This so-called reactor neutrino anomaly also strengthened the case
in favor of neutrino oscillations driven by 
$\Delta m^2 \sim 1$ eV$^2$. 
Recently, new model-independent hints in favor of SBL
$\bar\nu_e$ oscillations have emerged from the
reactor experiments NEOS~\cite{Ko:2016owz} and 
DANSS~\cite{DANSS:2017}. By performing the combined
analysis of the spectral ratios of these two experiments,
the authors of Ref.~\cite{Gariazzo:2018mwd}
have obtained an indication ($\sim$ 3.7$\sigma$) 
in favor of SBL $\bar\nu_e$ oscillations\footnote{However, there are
SBL experiments like CCFR~\cite{Stockdale:1984cg},
CDHSW~\cite{Dydak:1983zq}, and
SciBooNE-MiniBooNE~\cite{Mahn:2011ea,Cheng:2012yy}
that have observed null results while searching for muon flavor 
disappearance associated with this high oscillation 
frequency. The same is also true for the long-baseline 
experiment MINOS~\cite{MINOS:2016viw} 
and the atmospheric neutrino experiments 
Super-Kamiokande~\cite{Abe:2014gda} and IceCube~\cite{TheIceCube:2016oqi}.
Thus, there are tensions among various data sets, while trying to
fit all of them in a four-neutrino framework~\cite{Collin:2016aqd,Capozzi:2016vac,Gariazzo:2017fdh,Dentler:2017tkw,Dentler:2018sju}.
Therefore, the existence of a light sterile neutrino is 
still inconclusive.}
with $\Delta m^2 \approx$ $1.3$ eV$^2$.
A recent reanalysis of the ILL reactor data~\cite{Cogswell:2018auu}
also claims a $\sim 3\sigma$ evidence
for a light sterile neutrino with $\Delta m^2 \sim 1$ eV$^2$.

All these anomalies recorded at the SBL experiments 
cannot be explained with the help of three sub-eV
massive neutrinos $\nu_1$, $\nu_2$, $\nu_3$
and require the existence of a fourth mass eigenstate
$\nu_4$ at the eV scale.
This cannot couple to $W$ and $Z$ bosons due
to the LEP bounds~\cite{ALEPH:2005ab} on the number of the weakly 
interacting light neutrino state. Hence it has to be a
gauge singlet, and can reveal its presence 
only through its mixing with the active neutrinos.
This non-interacting neutrino is popularly known as
the ``sterile'' neutrino.
New more sensitive SBL experiments have been planned 
(see the reviews in 
Refs.~\cite{Gariazzo:2015rra,Giunti:2015wnd,Stanco:2016gnl,Fava:2016vas})
to test the existence of sterile neutrino by observing the
typical $L/E$ pattern in the oscillations driven by the new
mass-squared splitting $\Delta m^2_{41} \sim 1$ eV$^2$.
Apart from the SBL oscillation experiments, 
the light sterile neutrino can also have visible effects in
ongoing~\cite{Klop:2014ima,Agarwalla:2016mrc,Ghosh:2017atj} or 
upcoming~\cite{Gandhi:2015xza,Agarwalla:2016xlg,Agarwalla:2016fkh,Choubey:2017cba,Agarwalla:2018} 
long-baseline experiments.
Signals of a sterile neutrino may be observed at the existing
or planned multi-purpose water- or ice-based large detectors like
Super-Kamiokande~\cite{Abe:2014gda},
IceCube~\cite{TheIceCube:2016oqi,Razzaque:2011ab,Esmaili:2012nz,Esmaili:2013vza,Blennow:2018hto}, or
DeepCore~\cite{Razzaque:2012tp,Esmaili:2013cja}, or during the 
neutrino burst from a galactic supernova~\cite{Choubey:2007ga,Esmaili:2014gya}. 
The $\beta$-decay and 
neutrinoless double $\beta$-decay experiments 
can also feel the presence of a light sterile neutrino 
(see the review~\cite{Giunti:2015wnd} and the references therein).
The possible existence of a light sterile neutrino 
can also have profound implications in 
cosmology~\cite{Archidiacono:2013fha,Lesgourgues:2014zoa,Gariazzo:2015rra}.

The sterile neutrino is an elementary particle beyond the 
Standard Model (SM) and the possible discovery of a 
light sterile neutrino would prove that there is new physics 
beyond the SM at low-energies, which is completely orthogonal 
to new physics searches at high-energies at the 
Large Hadron Collider (LHC). There are several interesting 
motivations to search for a light sterile neutrino at mass
scales different than the eV scale. 
For instance, the possible existence of a super-light sterile 
neutrino~\cite{deHolanda:2003tx,deHolanda:2010am,Dev:2012bd,Liao:2014ola,Divari:2016jos}, 
which weakly mixes with the active neutrinos and 
has a mass very close to the active ones 
($\Delta m^2_{41}$  $\approx$ 10$^{-5}$ eV$^2$),
may explain the suppression of the upturn in the energy
spectrum of solar neutrino events below 
8 MeV~\cite{Abe:2016nxk,Aharmim:2011vm,Agostini:2017ixy}. 
A very light sterile neutrino at a mass scale smaller 
than 0.1 eV could affect the oscillations of reactor 
antineutrinos~\cite{Palazzo:2013bsa,Esmaili:2013yea,Girardi:2014wea,An:2014bik,An:2016luf}. 
A light sterile neutrino having a mass of a few eV 
can help in nucleosynthesis of heavy elements 
inside the supernova~\cite{Caldwell:1999zk,Tamborra:2011is,Wu:2013gxa}.
A relatively heavy sterile neutrino with mass around keV 
scale can act as warm dark matter in the context of the Neutrino
Minimal Standard Model 
($\nu$MSM)~\cite{Asaka:2005an,Asaka:2005pn,Boyarsky:2009ix,Adhikari:2016bei}. 
Therefore, it makes perfect sense to investigate
the presence of a light sterile neutrino over a wide 
range of mass scale.
Recently, the Daya Bay Collaboration also has
performed such a search and has obtained bounds on
the active-sterile mixing parameters in 
the mass range of $2 \times 10^{-4}$ eV$^2$ $\lesssim 
|\Delta m^2_{41}| \lesssim 0.3$ eV$^2$~\cite{An:2016luf}.

In this paper, we perform a detailed analysis 
of the prospects for the search for a light sterile
neutrino over a wide $\Delta m^2_{41}$ range, starting from $10^{-5}$ 
eV$^2$ and going all the way up to $10^2$ eV$^2$,
using 10 years of atmospheric neutrino data 
expected from the proposed 50 kt magnetized 
Iron Calorimeter (ICAL) detector~\cite{Kumar:2017sdq}
under the India-based Neutrino Observatory 
(INO) project~\cite{INO}. The magnetized INO-ICAL 
detector would detect the atmospheric $\nu_{\mu}$ 
and $\bar\nu_{\mu}$ separately over a wide range 
of energies and baselines. These features would help the 
ICAL detector to probe the presence of 
$\Delta m^2_{41}$ over a wide range and 
also to determine its sign.
The charged-current (CC) interactions of
$\nu_{\mu}$ and $\bar\nu_{\mu}$ inside the
ICAL detector will produce $\mu^{-}$ and $\mu^{+}$
particles, respectively. The ICAL detector would be able to
measure the energies and directions of these muons to a very good 
precision~\cite{Chatterjee:2014vta}. 
Moreover, the ICAL detector would be sensitive 
to neutrinos in the multi-GeV range, and the 
energies of hadron showers produced by 
the interactions of these multi-GeV neutrinos can also be 
well-estimated~\cite{Devi:2013wxa,Mohan:2014qua}.
While the main aim of this detector 
is to resolve the issue of neutrino mass 
ordering~\cite{Ghosh:2012px,Devi:2014yaa,Ajmi:2015uda,Kumar:2017sdq} 
and to improve the precision on atmospheric neutrino
mixing parameters~\cite{Thakore:2013xqa,Devi:2014yaa,Kaur:2014rxa,Mohan:2016gxm,Kumar:2017sdq},  
the above features can also be instrumental 
in the search for a light sterile neutrino.

A sensitivity study of ICAL to a sterile neutrino in the
  range $0.1$ eV$^2 \lesssim \dm_{41} \lesssim 10$ eV$^2$ has recently been
  performed~\cite{Behera:2016kwr}, where exclusion limits on the mixing
  parameters with 500 kt-yr exposure have been shown.
  In this paper, we not only extend this analysis to much lower values
  of $\dm_{41}$ (all the way down to $10^{-5}$ eV$^2$),
  but also point out for the first time that in the $\dm_{41}$ range of
  $(0.5 - 5) \times 10^{-3}$ eV$^2$,  
  ICAL has a far greater sensitivity to the magnitude as well as
  sign of $\dm_{41}$.
  We also explore the impact of hadron energy calibration, the actual
  value of $|U_{e4}|$, up-going vs. down-going events, 
  spectral information, and the signs of $\Delta m^2_{31}$ 
  as well as $\Delta m^2_{41}$, on the sensitivity of ICAL to a sterile
  neutrino.
  In addition, in the scenario where a light sterile neutrino exists in
  Nature, we study the role of ICAL, in particular its muon charge
  identification capability, in identifying the mass ordering
  configuration of the four-neutrino spectrum.

We plan this article in the following fashion.
We start Sec.~\ref{framework} with a 
discussion on different mass ordering schemes
and configurations which are possible in the 
four-neutrino (4$\nu$) framework. Then we describe the 
lepton mixing matrix in the 4$\nu$ scheme and
mention the present constraints that we have 
on $|U_{e4}|^2$, $|U_{\mu 4}|^2$, and $|U_{\tau 4}|^2$
from various oscillation experiments. We further elaborate
on the Earth matter effects in the presence of three active
neutrinos and one light sterile neutrino. We end this section
by drawing the neutrino oscillograms in 
$E_\nu$--$\cos\theta_\nu$ plane in the 4$\nu$ paradigm.
In Sec.~\ref{sec:analysis}, we mention our analysis
procedure and present the expected
event spectra at the ICAL detector as a function of
$E_{\mu}$ and $\cos\theta_{\mu}$ for several
benchmark choices of active-sterile oscillation parameters.
We identify the regions in $E_{\mu}$ and 
$\cos\theta_{\mu}$ plane which give significant 
contributions in constraining active-sterile oscillations.
In Sec.~\ref{sec:constraints}, we determine the 
parameter space in the $|\Delta m^2_{41}|$--$|U_{\mu 4}|^2$
plane that can be excluded by the ICAL data if there
is no light sterile neutrino in Nature.
We also study the dependence of these constraints on 
$|U_{e4}|$, mass ordering schemes, as well as
energy and direction measurements.
In Sec.~\ref{sec:discovery}, we assume
that there is a light sterile neutrino in Nature and then
address several interesting issues, like the precision 
in the determination of $\Delta m^2_{41}$ and 
the chances of measuring its sign at
the INO-ICAL atmospheric neutrino experiment. 
Finally in Sec.~\ref{sec:conclusions}, we summarize 
and draw our conclusions.

\section{Neutrino mixing and oscillations in four flavors}
\label{framework}

\subsection{Mass ordering schemes and configurations}

\begin{figure}
\centering
\includegraphics[width=0.49\textwidth]{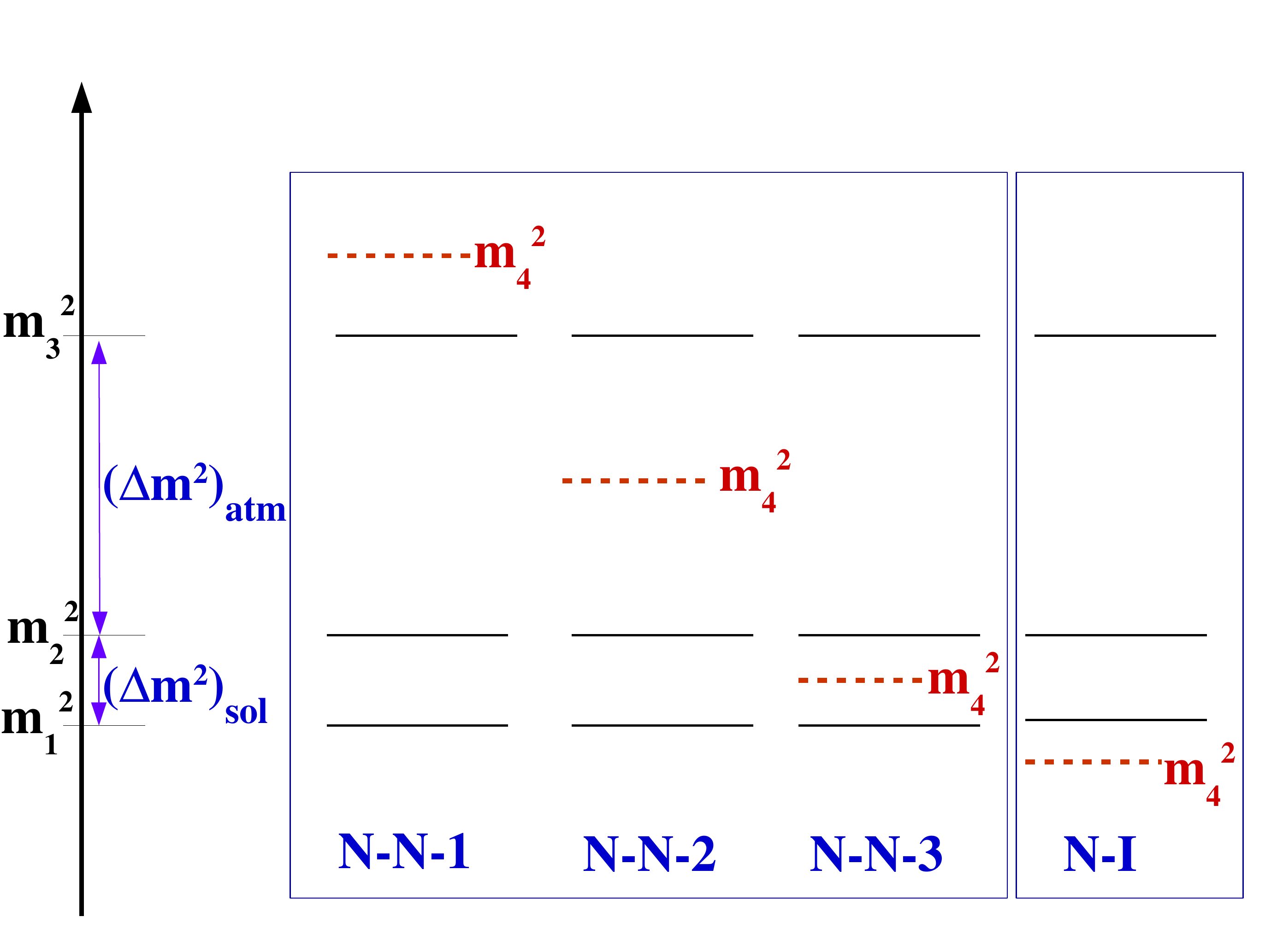}
\includegraphics[width=0.49\textwidth]{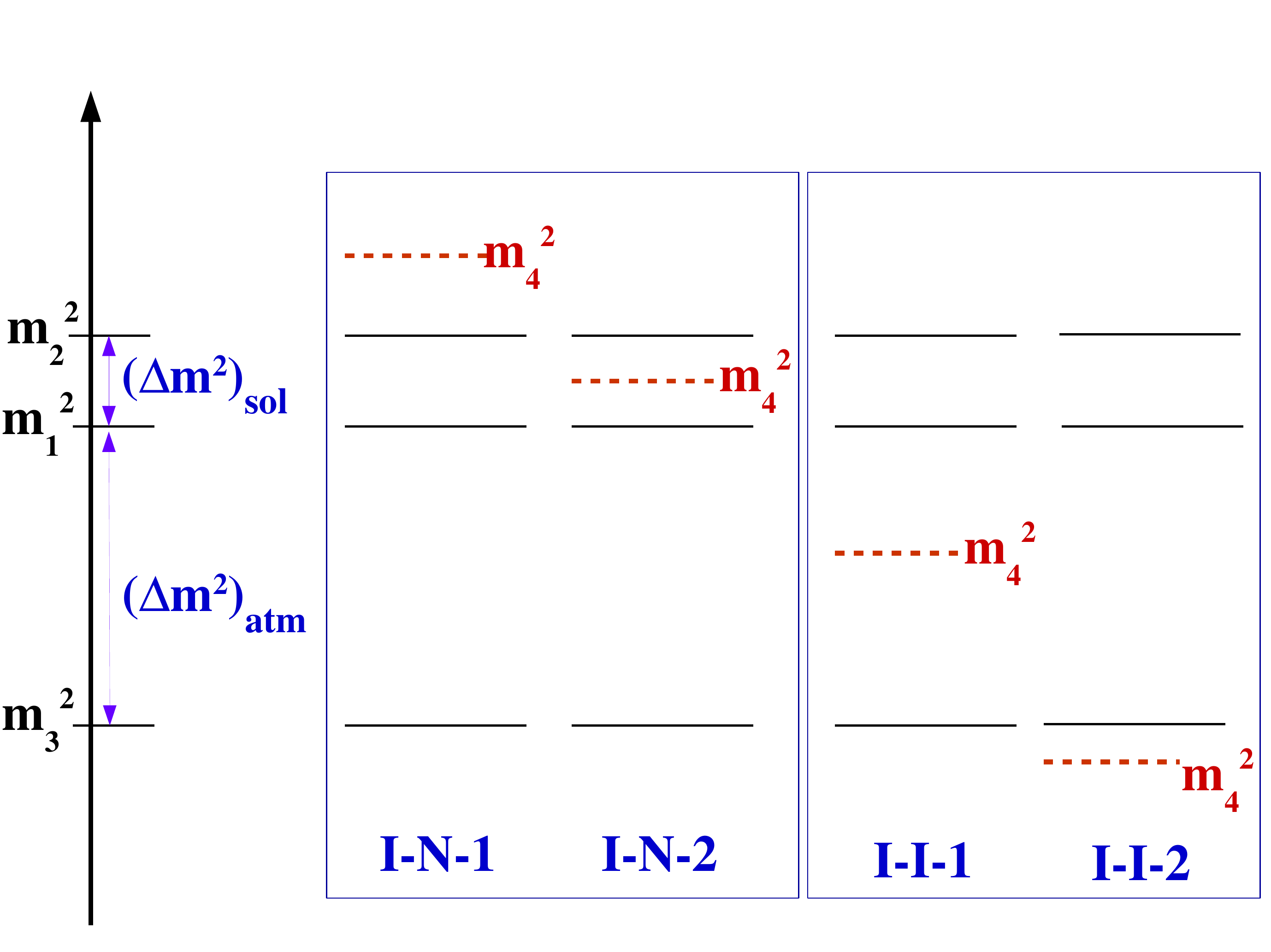}  
\caption{Graphical representations of the mass ordering configurations
  given in Table~\ref{tab:config}.}
\label{fig:config}
\end{figure}

The three active neutrinos $\nu_{e,\mu,\tau}$ and the sterile neutrino $\nu_s$
give rise to four mass eigenstates, $\nu_{1,2,3,4}$. Here $\nu_{1,2,3}$ are
dominated by active flavors, in the decreasing order of electron flavor
fraction, while $\nu_4$ is dominated by the sterile neutrino.
The mass eigenstates may be arranged, according to their relative
mass squared values, in eight possible configurations that are allowed
by the current data. We label these configurations initially according to
the signs of $\dm_{31}$ and $\dm_{41}$ (N-N, N-I, I-N and I-I,
as shown in the top row of Table~\ref{tab:config}),
and further by
the relative position of $\nu_4$ on the $m_i^2$-scale, which will
depend on the signs of $\dm_{42}$ and $\dm_{43}$, as indicated in the
lower rows of Table~\ref{tab:config}.
The configurations have also been shown
graphically in Fig.~\ref{fig:config}.

\begin{table}[t]
\begin{center}
  \begin{tabular}{|l|c|c|c|c||c|c|c|c|}
    \hline 
  Ordering scheme & \multicolumn{3}{c|}{N-N} & N-I &
  \multicolumn{2}{c}{I-N} & \multicolumn{2}{|c|}{I-I} \\
  \hline
  Sign of $\dm_{31}$ & \multicolumn{3}{c|}{+} & + &
  \multicolumn{2}{c}{-} &\multicolumn{2}{|c|}{-} \\
  Sign of $\dm_{41}$ & \multicolumn{3}{c|}{+} & - &
  \multicolumn{2}{c}{+} &\multicolumn{2}{|c|}{-} \\
  \hline
  Sign of $\dm_{42}$ & + & + & - & - & + & - & - & - \\
  Sign of $\dm_{43}$ & + & - & - & - & + & + & + & - \\
  \hline
  Configuration & N-N-1 & N-N-2 & N-N-3 & N-I & I-N-1 & I-N-2 & I-I-1 & I-I-2 \\
  \hline
  \end{tabular}
\end{center}
\caption{All possible neutrino mass ordering schemes and configurations in the
  four-neutrino framework.}
\label{tab:config}
\end{table}

In this paper, we shall focus on exploring these configurations 
by observing atmospheric neutrinos over a wide range of energies 
and baselines at INO-ICAL. Some of these configurations, 
viz. N-N-1 and I-N-1, correspond to the sterile neutrino solution 
required to satisfy the SBL anomalies discussed in the previous
section, in the so-called 3+1 
scheme~\cite{Goswami:1995yq,Okada:1996kw,Bilenky:1996rw,Bilenky:1999ny,Maltoni:2004ei}, 
while still being consistent with the cosmological bounds 
on the sum of neutrino masses~\cite{Ade:2015xua}.
Some others, viz. N-N-3 and I-N-2, may be relevant for a super-light
sterile neutrino that has been 
proposed~\cite{deHolanda:2003tx,deHolanda:2010am,Liao:2014ola,Divari:2016jos}
to explain the non-observation of the upturn in the low energy 
solar data~\cite{Abe:2016nxk,Aharmim:2011vm,Agostini:2017ixy}. 
However, in this work, we shall not be looking for a solution 
to any of these anomalies, rather we shall perform an 
agnostic search for a sterile neutrino that
may belong to any of these configurations.
Note that the configurations N-I and I-I-2 
will be disfavoured by the cosmological 
bounds~\cite{Ade:2015xua}, unless the 
magnitude of $\dm_{41}$ is very small. 

\subsection{The Lepton Mixing Matrix}

In the 4$\nu$ scheme, the four flavor eigenstates
($\nu_e,\nu_\mu,\nu_\tau, \nu_s$)
are linked to the mass eigenstates ($\nu_1,\nu_2,\nu_3,\nu_4$) via a $4\times4$
unitary lepton mixing matrix, which has the following form:
\begin{equation}
U \equiv \left( \begin{array}{cccc} 
U_{e1} & U_{e2} & U_{e3} & U_{e4} \\
U_{\mu 1} & U_{\mu 2} & U_{\mu 3} & U_{\mu 4} \\
U_{\tau 1} & U_{\tau 2} & U_{\tau 3} & U_{\tau 4} \\
U_{s1} & U_{s2} & U_{s3} & U_{s4} \\
\end{array} \right) \; .
\end{equation}
Here, we expect $|U_{e4}|^2, |U_{\mu 4}|^2, |U_{\tau 4}|^2 \ll 1$, while 
$|U_{s4}|^2 \approx 1$~\cite{Collin:2016aqd,Capozzi:2016vac,Gariazzo:2017fdh,Dentler:2017tkw,Dentler:2018sju}.
There are constraints on active-sterile mixing parameters from 
atmospheric, long-baseline, as well as reactor experiments.

The Super-Kamiokande Collaboration limits $|U_{\mu 4}|^2$ to less than
0.04 for $\dm_{41} > 0.1$ eV$^2$ at 90\% C.L. in N-N-1 configuration 
using 4438 live-days of atmospheric neutrino data and assuming 
$|U_{e4}|^2 = 0$~\cite{Abe:2014gda}. The IceCube neutrino telescope
at the South Pole searches for active-sterile oscillation in their measured
atmospheric muon neutrino spectrum as a function of zenith angle and 
energy in the 320 GeV -- 20 TeV range~\cite{TheIceCube:2016oqi}.
The muon antineutrinos in their data sample do not experience a 
strong MSW resonance effect, which allows them to obtain a
bound on the active-sterile mixing. As an example, 
for $\Delta m^2_{41} \approx 1.75$ eV$^2$ (N-N-1 configuration), which is the  
best-fit value of $\Delta m^2_{41}$ according to Ref.~\cite{Collin:2016aqd},
the IceCube data give a bound of $\sin^22\theta_{24} \lesssim 0.06$
at the 90\% C.L.\footnote{
In the short-baseline approximation~\cite{Bilenky:1996rw}, the effective
mixing angle $\sin^22\theta_{24} \equiv 4|U_{\mu4}|^2(1-|U_{\mu4}|^2) 
\approx 4|U_{\mu4}|^2$, so this bound corresponds to 
$|U_{\mu 4}|^2 \lesssim 0.015$.}.
In the same $\dm_{41}$ range, 
the MINOS Collaboration places a new upper limit of 
$\sin^2\theta_{24} < 0.03$ at 90\% C.L. using their long-baseline data,
suggesting that
$|U_{\mu 4}|^2 \lesssim 0.03$~\cite{MINOS:2016viw,Adamson:2016jku}.

The combined analysis of the available data from the reactor experiments  
Daya Bay and Bugey-3 provides a new upper limit of
$\sin^22\theta_{14} \lesssim 0.06$
at 90\% C.L. \footnote{In the 
short-baseline approximation~\cite{Bilenky:1996rw}, the effective
mixing angle, $\sin^22\theta_{14} \equiv 4|U_{e4}|^2(1-|U_{e4}|^2) 
\approx 4|U_{e4}|^2$, so this bound corresponds to 
$|U_{e4}|^2 \lesssim 0.015$. Occasionally, $\sin^22\theta_{14}$ 
is also denoted as $\sin^22\theta_{ee}$.} around 
$\Delta m^2_{41} \approx 1.75$ eV$^2$~\cite{Adamson:2016jku}.
Recently, the reactor antineutrino experiments NEOS~\cite{Ko:2016owz} 
and DANSS~\cite{DANSS:2017} have provided new hints in favor
of short-baseline $\bar\nu_e$ oscillations, which are model independent
in the sense that these indications are not dependent on the precise 
estimate of the reactor $\bar\nu_e$ fluxes. In N-N-1 configuration,
the combined analysis of the DANSS and NEOS spectral ratios 
predict a narrow-$\Delta m^2_{41}$ island at 
$\Delta m^2_{41} \approx 1.75$ eV$^2$ with 
$\sin^22\theta_{ee} = 0.049 \pm 0.023$ at 
2$\sigma$~\cite{Gariazzo:2018mwd}, which means that 
the best-fit value of $|U_{e4}|^2$ is around 0.012.
Another recent study on the same topic can be found in
Ref.~\cite{Dentler:2017tkw}.

As far as the constraint on $|U_{\tau 4}|^2$ is concerned, the Super-Kamiokande 
experiment places an upper limit of 0.18 on this parameter at 90\% C.L.,
for $\Delta m^2_{41} > 1$ eV$^2$~\cite{Abe:2014gda}.
The IceCube DeepCore Collaboration sets upper 
bounds of $|U_{\mu 4}|^2 < 0.11$ and 
$|U_{\tau 4}|^2 < 0.15$ at 90\% C.L. for 
$\Delta m^2_{41} \sim 1$ eV$^2$, using their three
years of atmospheric neutrino data in the range 
10 to 60 GeV~\cite{Aartsen:2017bap}.
One can test the mixing between sterile and
tau neutrino by looking for a depletion in the 
neutral-current event rates at the far detector
of a long-baseline setup. Both the 
MINOS~\cite{Adamson:2011ku} and 
NO$\nu$A~\cite{Adamson:2017zcg} experiments
place competitive constraints on mixing
between sterile and tau neutrinos using this approach.
According to the global fit study performed 
in Ref.~\cite{Gariazzo:2017fdh}, 
$|U_{\tau 4}|^2 \lesssim 0.014$ at 90\% C.L.
for any value of $\Delta m^2_{41}$ 
in N-N-1 configuration. The authors of 
Ref.~\cite{Collin:2016aqd} obtain similar
bounds on $|U_{\tau 4}|^2$ around 
$\Delta m^2_{41} \approx 6$ eV$^2$.

The INO-ICAL experiment will probe a combination of the 
$\nu_e \to \nu_\mu$ and $\nu_\mu \to \nu_\mu$ oscillation channels, 
mainly via the charged-current (CC) interactions that produce muons. 
The active-sterile mixing elements of relevance here are therefore 
$U_{e4}$ and $U_{\mu 4}$ (and to some extent $U_{\tau 4}$, due to 
Earth matter effects). In order to illustrate the impact of active-sterile 
oscillations for different choices of $\dm_{41}$,
we shall be using the benchmark 
values $|U_{e4}|^2 = 0.025$, $|U_{\mu 4}|^2 = 0.05$, and 
$|U_{\tau 4}|^2 = 0$ while showing our oscillograms and event 
plots. For most of our sensitivity plots, we shall take $|U_{e4}|^2=0$, 
unless otherwise mentioned. We shall also demonstrate later that 
$|U_{e4}|^2 = 0$ yields the most conservative bounds 
in the $(\dm_{41}, |U_{\mu 4}|^2)$ parameter space.
As far as the three-neutrino oscillation parameters are concerned,
Table~\ref{tab:bench} shows the benchmark values that we consider 
in this work.

\begin{table}[t]
\centering
\begin{tabular}{|l|l|l|l|l|l|l|}
\hline
Parameter & $\sin^22\theta_{12}$ & $\sin^2\theta_{23}$ & $\sin^22\theta_{13}$ & 
$\Delta m_{21}^2 (\rm eV^2)$ & $|\Delta m_{31}^2| (\rm eV^2)$ & $\dcp$ \\
\hline
Value & 0.84 & 0.5 & 0.1 & 7.5 $\times$ $10^{-5}$ 
& $2.4 \times$ $10^{-3}$ & 0$^\circ$ \\
\hline
\end{tabular}
\caption{Benchmark values of the three-flavor neutrino oscillation parameters 
considered in this work.}
\label{tab:bench}
\end{table}

\subsection{Earth matter effects}

The probability for a neutrino produced 
with flavor $\alpha$ and energy $E$ to be detected as a neutrino of flavor 
$\beta$ after traveling a distance $L$ in vacuum can be written
in terms of the leptonic mixing matrix elements as
\cite{Barger:1999hi,Kayser:2002qs}:
\begin{eqnarray}
P(\nu_{\alpha}\to\nu_{\beta}) = & \delta_{\alpha\beta}
- 4\sum_{i>j} {\rm Re}(U^{\ast}_{\alpha i}
U_{\beta i}U_{\alpha j}U^{\ast}_{\beta j})\sin^2 \Delta_{ij} \nonumber \\
& \phantom{space}
+ 2\sum_{i>j} {\rm Im}(U^{\ast}_{\alpha i}U_{\beta i}U_{\alpha j}U^{\ast}_{\beta j})
\sin 2 \Delta_{ij} \,,
\label{eq:oscprob} 
\end{eqnarray}
where $\Delta_{ij}\equiv \Delta m^2_{ij}L/(4E)$.
For antineutrinos, the oscillation probability follows Eq.~\ref{eq:oscprob} 
with the replacement of $U$ with its complex-conjugate matrix.
While propagating through the Earth, the elements of the leptonic mixing
matrix change, since the forward scattering of neutrinos on the
Earth matter gives rise to effective matter potentials 
\begin{equation}
\begin{array}{ccll}
\phantom{V_{es} = } V_{es}  & = & \sqrt{2} G_F (N_e - N_n/2)  & \qquad 
\mbox{between $\nu_e$ and $\nu_s$} \; , \\
V_{\mu s} = V_{\tau s}  & = & - \sqrt{2} G_F N_n/2  & \qquad 
\mbox{between $\nu_{\mu/\tau}$ and $\nu_s$} \; , \\
\end{array}
\end{equation}
in the flavor basis. Here $N_e$ is the electron number density and $N_n$
is the neutron number density inside the Earth.
These modified elements need to be used while computing the neutrino
oscillation probabilities through layers of different densities
inside the Earth.

As neutrinos start crossing longer distances through the
Earth, the Earth matter effects start becoming important, as
the relative matter potentials $V_{e s}$ and $V_{\mu s}$ 
are $\sim \dm_{31}/(2E)$ for atmospheric neutrinos (e.g. for
$E=5$ GeV and $L=5000$ km).
As a consequence, for $\dm_{41} \sim \dm_{31}$, the Earth matter
affects active-sterile conversion probabilities significantly.
In this $\dm_{41}$ region, the similar magnitudes $\Delta m^2_{41}$ and
$\Delta m^2_{31}$ also lead to an interference between the oscillation
frequencies governed by them,
which will manifest itself in the sensitivity of the ICAL
experiment in the $\dm_{41}$--$|U_{\mu 4}|^2$ plane, 
as we shall see later in our results. 
For $\dm_{41} \gg \dm_{31}$, which is the parameter space usually
focused on 
(since it is relevant for explaining the LSND anomaly), matter effects
are not very important. However for $\dm_{41} \ll \dm_{31}$, a region of
the parameter space that we specially explore, the Earth matter effects
would play a major role.

In order to compute the neutrino conversion probabilities in the
presence of Earth matter numerically, in both the three-flavor and
four-flavor scenarios, we use the GLoBES
software~\cite{Huber:2004ka,Huber:2007ji} along with its new physics tools.
To take into account the Earth matter effects, we take the
Preliminary Reference Earth Model (PREM) profile for the density of
the Earth \cite{PREM:1981}, with five density steps (to keep the computation
time manageable).
For the down-going neutrinos ($L \lesssim 450$ km), we take all the
neutrinos to be produced at an uniform height of 15 km, and consider
them to be travelling through vacuum.
For the upward-going neutrinos ($L \gtrsim 450$ km), we neglect the
height of the atmosphere, which is small compared to the total distance
travelled. We have checked that these approximations
do not affect the oscillation probabilities, and hence the event
distribution in energy and zenith angle, to any appreciable extent.

\subsection{Oscillograms in $E_\nu$--$\cos\theta_\nu$ plane}
\label{sec:oscillogram}

In order to get an idea of the impact of active-sterile mixing on the
$\nu_\mu \to \nu_\mu$ survival probability $P_{\mu \mu}$ and the
$\nu_e \to \nu_\mu$ conversion probability $P_{e \mu}$, we define
the quantities
\begin{equation}
\Delta P_{\mu\mu} \equiv P_{\mu\mu} \mbox{ (4f) } -P_{\mu\mu} \mbox{ (3f) }
\quad , \qquad \mbox{and} \qquad 
\Delta P_{e \mu} \equiv P_{e \mu} \mbox{ (4f) } 
- P_{e\mu} \mbox{ (3f) } \; ,
\end{equation}
where ``(4f)'' and ``(3f)'' denote the quantities calculated in the
4-flavour and 3-flavour mixing scenario, respectively.
In Fig.~\ref{oscillogram:pmumu-pemu}, we plot these quantities in
the $E_\nu$--$\cos\theta_\nu$ plane, where $\theta_\nu$ is the zenith angle
and $E_\nu$ is the energy of the neutrino.
We present our results for two different values of $\dm_{41}$,
viz. $\dm_{41} = 1$ eV$^2$ (corresponding to the N-N-1 configuration), and
$\dm_{41} = 10^{-3}$ eV$^2$ (corresponding to the N-N-2 configuration).
In the former scenario, we average over the fast oscillations 
when neutrinos travel large distances through the Earth. 
From the figure, the following observations may be made:

\begin{figure}[t]
\centering
\includegraphics[width=0.49\textwidth]{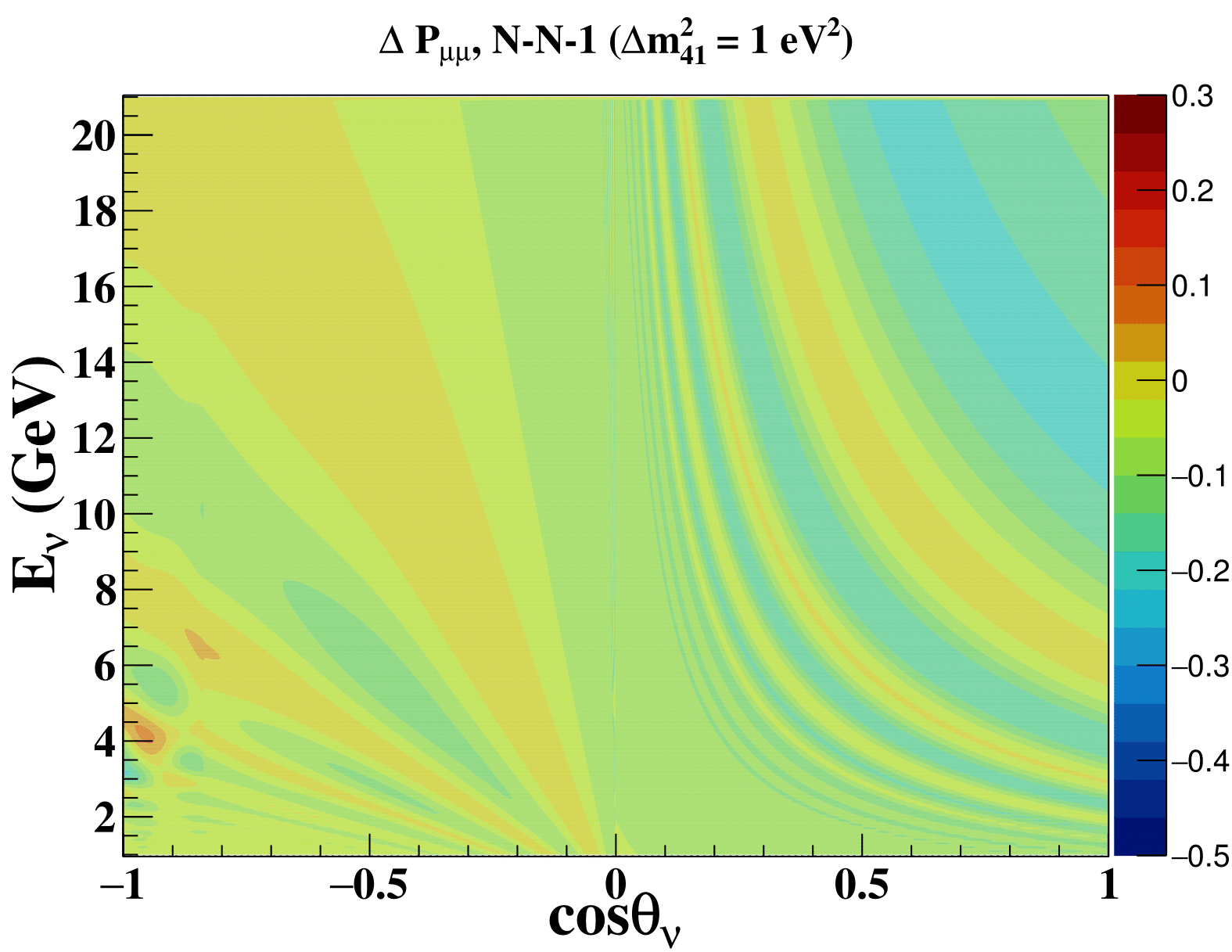}
\includegraphics[width=0.49\textwidth]{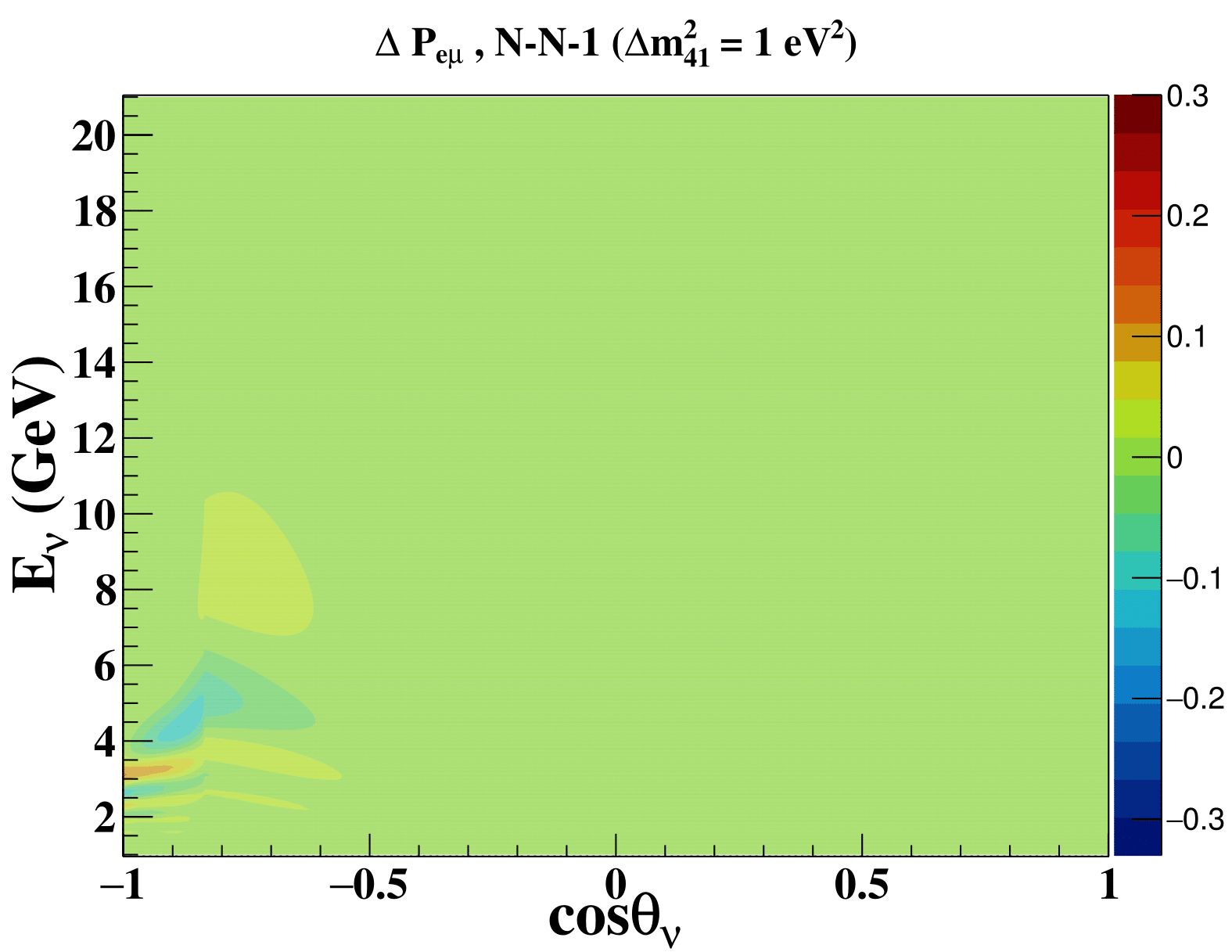}
\includegraphics[width=0.49\textwidth]{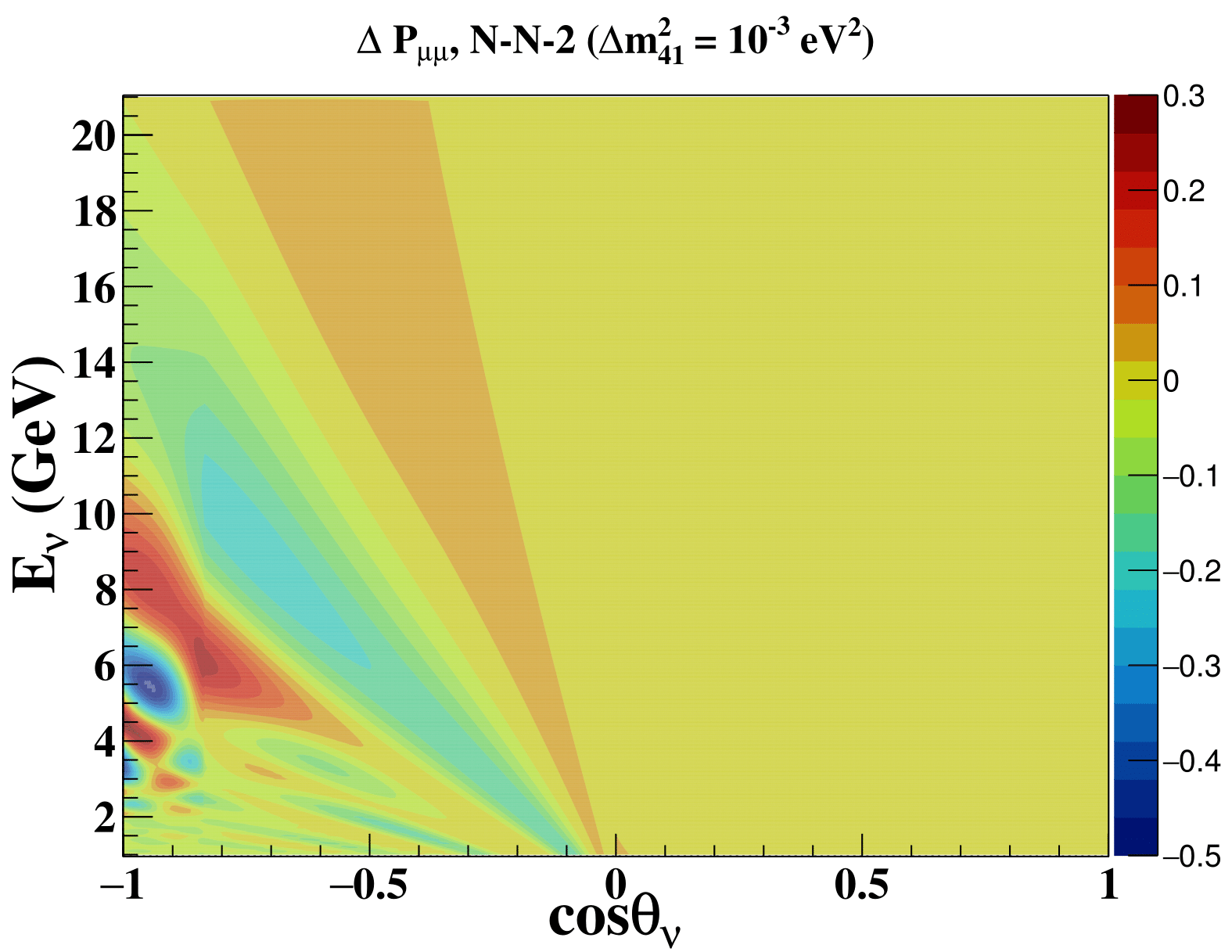}
\includegraphics[width=0.49\textwidth]{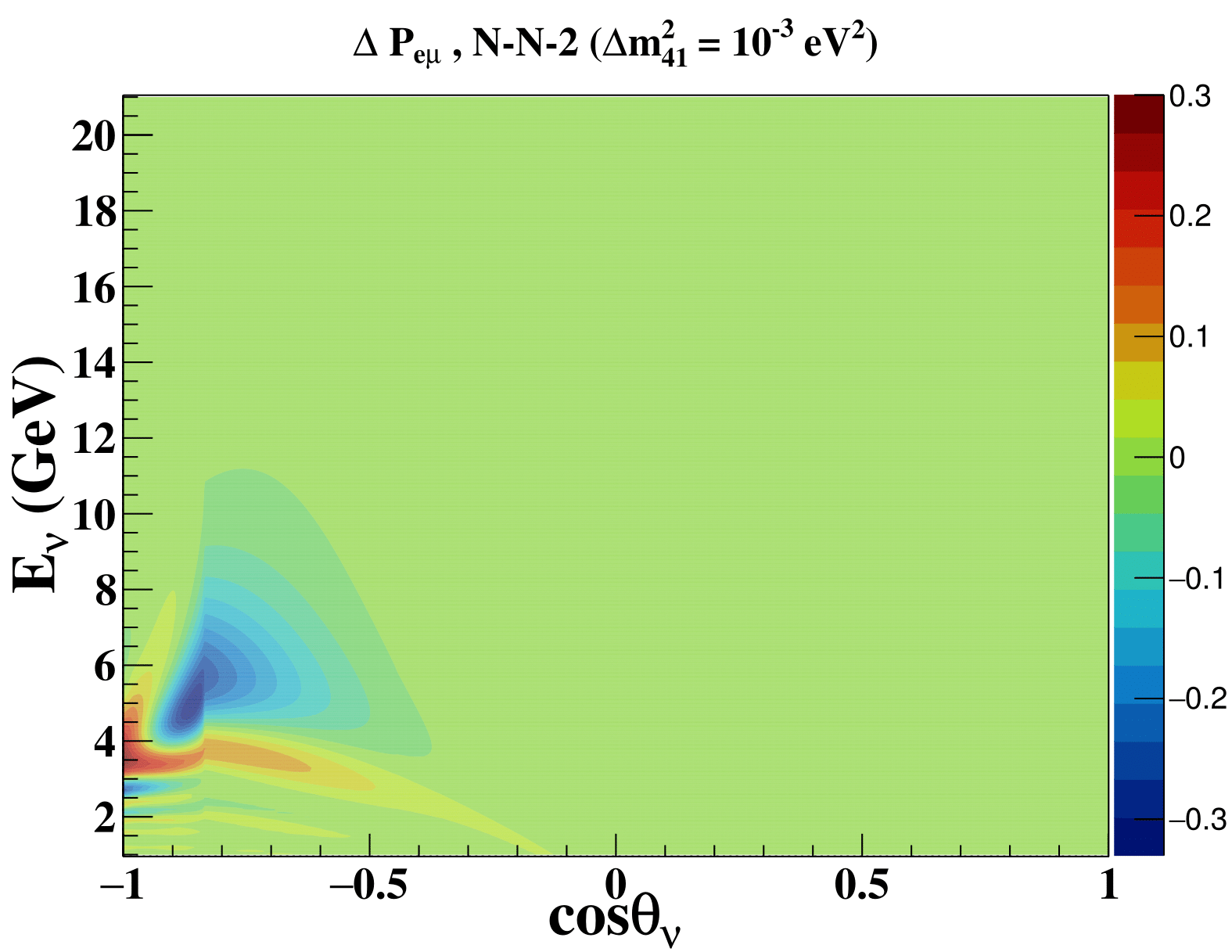}
\mycaption{Oscillograms for $\Delta P_{\mu\mu}$ (left panels) and 
$\Delta P_{e \mu}$ (right panels), and for $\Delta m^2_{41}=1$ eV$^2$ (top panels)
and $\Delta m^2_{41}=10^{-3}$ eV$^2$ (bottom panels), after performing 
the averaging over the fast oscillations. We have used 
$|U_{e 4}|^2 = 0.0025$, $|U_{\mu 4}|^2 = 0.05$, and $|U_{\tau 4|^2}=0$.
The hierarchy in the active sector has been taken to be normal
($\Delta m^2_{31} > 0$).}
\label{oscillogram:pmumu-pemu}
\end{figure}

\begin{itemize}

\item Let us first see the impact of the active-sterile mixing in
  the $\nu_\mu \rightarrow \nu_\mu$ survival channel, which plays an
  important role in atmospheric neutrino oscillations.
  The effect of sterile neutrino mixing on $P_{\mu\mu}$ is clearly significant.
  For $\Delta m^2_{41} = 1$ eV$^2$, the effect is observed over a wide region in
  $E_\nu$ and $\cos\theta_\nu$. For $\Delta m^2_{41} = 10^{-3}$ eV$^2$,
  the impact is mostly confined to the upward going neutrinos,
  while its magnitude is significant around 
  $E_\nu \in [3, 10]$ GeV and $\cos\theta_\nu \in [-1 , - 0.7]$. 

\item  
  While fast oscillations for the high $\Delta m^2_{41} \sim 1$ eV$^2$
  will get averaged out due to the finite energy and angular resolutions
  of the detector,   
  at $\Delta m^2_{41} \sim 10^{-3}$ eV$^2$ the oscillation probability
  dependence on energy and direction of neutrinos would be more clearly
  resolvable. Hence we expect that the spectral and angular information
  should play a role in identifying signatures of sterile neutrinos at
  low $\Delta m^2_{41}$ values .

\item The impact of $4\nu$ mixing on $P_{e\mu}$ is confined to a narrow region 
  to the upward-going neutrinos, and it is more prominent at the lower 
  $\Delta m^2_{41}$. This would indicate a significant contribution of
  the matter effects due to the Earth, which influence the upward-going
  neutrinos. 

\end{itemize}

The much higher effects of active-sterile mixing on $P_{\mu\mu}$, combined
with the much higher value of $P_{\mu\mu}$ as compared to $P_{e \mu}$,
and the higher fraction of $\nu_\mu$ in the atmospheric neutrino flux,
indicates that the $\nu_\mu \to \nu_\mu$ conversions will play a dominant
role in the sensitivity of atmospheric neutrino detectors to sterile
neutrinos. The corresponding antineutrino oscillation probability plots
show the same features.
In fact, the oscillograms at the higher $\Delta m^2_{41}$
values are virtually identical with those for neutrinos, shown
in the top panels of Fig.~\ref{oscillogram:pmumu-pemu}.
Therefore\footnote{Note that there are small differences between neutrino and
  antineutrino oscillograms for $\Delta P_{e\mu}$ at lower $\Delta m^2$
  for large propagation distances through the Earth matter, however
  these have a very small impact on our analysis.}, 
we should be able to interpret most of our results 
based on $P_{\mu\mu}$.

The oscillograms of $P_{\mu\mu}$ at low $\Delta m^2_{41}$ indicate the presence
of significant Earth matter effects. This points to the possible sensitivity
of the data to the sign of $\Delta m^2_{41}$, a point we shall explore further
in Sec.~\ref{sec:sign}.

\section{Event spectra at ICAL and identifying crucial $(E_\mu, \cos\theta_\mu)$ bins}
\label{sec:analysis}

\subsection{Event distributions in energy and zenith angle}
\label{sec:events}

In order to perform the simulations of the events in INO-ICAL,
we generate the events using the same procedure as described in
\cite{Ghosh:2012px,Devi:2014yaa,Kumar:2017sdq}.
The NUANCE event generator \cite{Casper:2002sd} is used to generate
unoscillated atmospheric neutrino events for 1000 years exposure of 
the 50 kt ICAL, which are later rescaled to the relevant exposure. 
The oscillations are implemented with the four-flavor probabilities in matter, 
using the reweighting algorithm, as described in \cite{Ghosh:2012px}.
The energy and direction resolutions of muons and hadrons are used, 
as per the latest results of the ICAL Collaboration \cite{Chatterjee:2014vta,
Devi:2013wxa}. We estimate the physics reach of ICAL in probing
active--sterile oscillation parameters by combining the muon momentum
information ($E_\mu$, $\cos\theta_\mu$) and the hadron energy information
($E'_{\rm had} \equiv E_\nu - E_\mu$) on an event-by-event basis.
For this work, it is assumed that the muon and hadron hits
can be separated with 100\% efficiency, and that the background and noise
are negligible.
The events are binned into 10 uniform $E_\mu$ bins in the range [1,11] GeV,
and 20 uniform $\cos\theta_{\mu}$ bins in the range $[-1,1]$.
The hadrons are binned in 4 $E'_{\rm had}$ bins: 1--2 GeV, 2--4 GeV, 
4--7 GeV and 7--11 GeV, as in~\cite{Devi:2014yaa}.

\begin{table}[t]
\centering
\begin{tabular}{|l|c|c|c|c|c|c|}
\hline
Case ($\Delta m^2_{41}$)  & Total $\mu^-$ & Total $\mu^+$ & Down $\mu^-$ & Down $\mu^+$ & Up $\mu^-$ & Up $\mu^+$ \\
\hline
3f  & 4736 & 2070 & 3108 & 1361 & 1628 & 709 \\
\hline
4f ($1$ eV$^2$), N-N-1  & 4296 & 1874 & 2817 & 1234 & 1479 & 641 \\
\hline
4f ($10^{-3}$ eV$^2$), N-N-2 & 4612 & 1984 & 3107 & 1361 & 1505 & 623 \\
\hline
4f ($10^{-5}$ eV$^2$), N-N-3 & 4664 & 2046 & 3108 & 1361 & 1557 & 685\\
\hline
\end{tabular}
\mycaption{Number of events for 500 kt-yr exposure of ICAL,
for $E_\mu \in [1,11]$ GeV and NH in the active sector. 
We have taken $|U_{e4}|^2=0.025$, $|U_{\mu 4}|^2=0.05$, and $|U_{\tau 4}|^2=0$.
The information on hadron energy is not used.
For sterile neutrinos, the results have been shown for three different  
$\Delta m^2_{41}$ values, corresponding to the N-N-1, N-N-2 and
N-N-3 configurations, respectively.
}
\label{tab:CC-events1}
\end{table}

Table~\ref{tab:CC-events1} gives the number of $\mu^{-}$ and $\mu^{+}$
events, in 3$\nu$ case and with three benchmark values of
$\Delta m^2_{41}$ for 4$\nu$ configurations. 
The hierarchy in the active sector has been taken to be normal
($\Delta m^2_{31} >0$). 
It may be observed that for $\Delta m^2_{41} = 1$ eV$^2$ (N-N-1 configuration),
the number of events with $|U_{\mu 4}|^2=0.05$ is about 10\% less than
the number with vanishing $|U_{\mu 4}|$, as expected 
after averaging over fast $\Delta_{41}$ oscillations.
For $\Delta m^2_{41} = 10^{-3}$ eV$^2$ (N-N-2 configuration),
the fractional difference in the total number of events with finite
vs. vanishing sterile mixing is rather small.
However Up/Down ratios of the $\mu^{-}$ and
$\mu^{+}$ events are different in these two scenarios, since
for these $\Delta m^2_{41}$ values, the active-sterile oscillation
probability will have a nontrivial zenith angle dependence.
Thus for low $\dm_{41}$, the information on angular distribution 
of these events may be useful.

  The last row in Table~\ref{tab:CC-events1} shows the scenario where
  $\dm_{41}$ is extremely small, $\dm_{41} = 10^{-5}$ eV$^2$ 
  (N-N-3 configuration). Note that
  even at this extremely low $\dm_{41}$ value, the effect of oscillations
  is still visible through the loss in number of events as well as
  the difference in the ratio of Up/Down events.
  This may be attributed
  to the fact that even when $\dm_{41} \to 0$, active-sterile oscillations
  may still occur through the frequencies corresponding to
  $|\dm_{42}| \sim 10^{-4}$ eV$^2$ and $|\dm_{43}| \sim 2.4 \times 10^{-3}$ eV$^2$.
  As a result, the $\dm_{41} \to 0$ limit in the 4$\nu$ mixing case
  does not correspond to the decoupling of sterile neutrinos, and
  limits on active-sterile mixing may still be obtained in this case.
  In our analysis, we go down to values of $\dm_{41}$ as low as
  $10^{-5}$ eV$^2$.

\begin{figure}[t]
\centering
\includegraphics[width=0.49\textwidth]{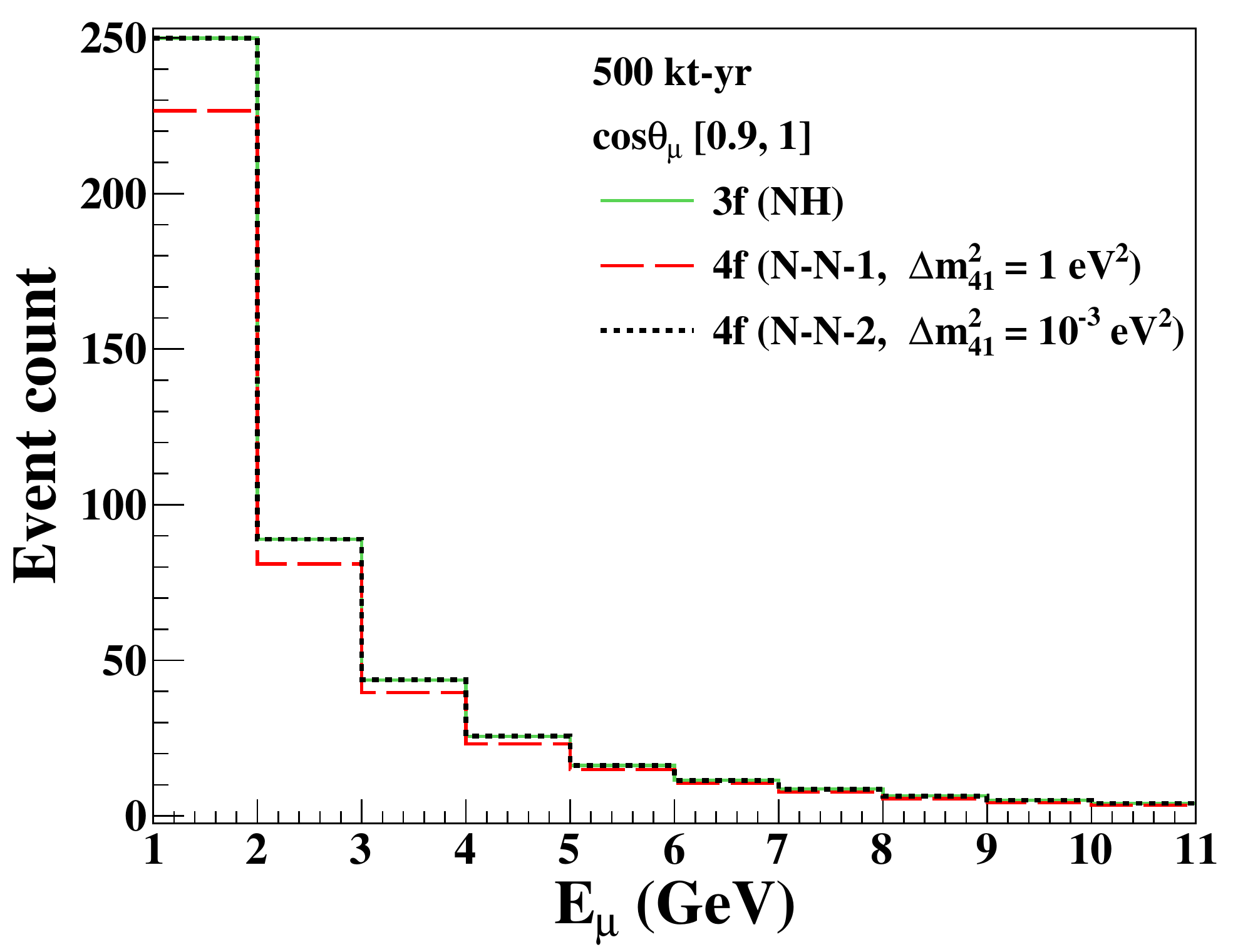}
\includegraphics[width=0.49\textwidth]{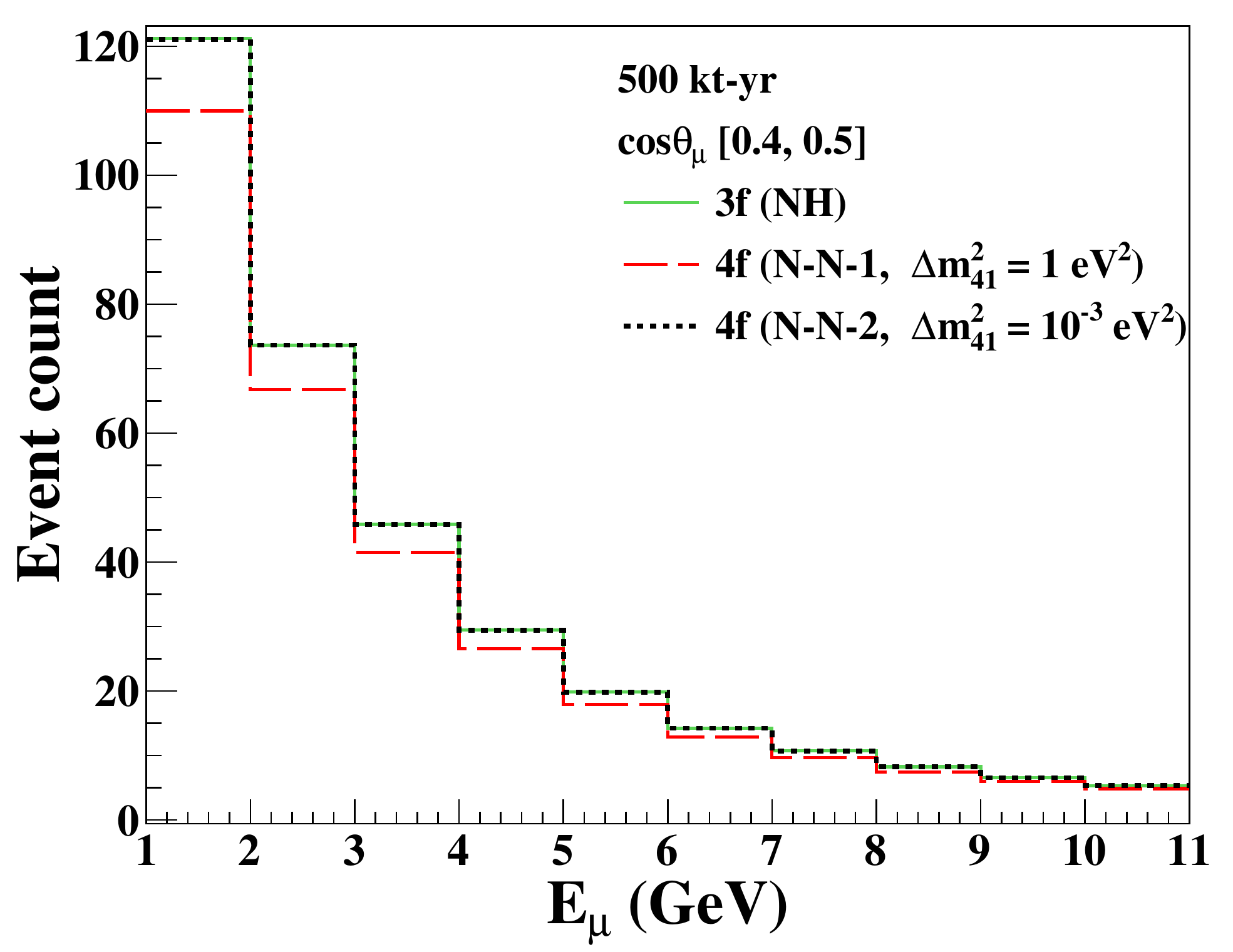}
\mycaption{$\mu^-$ event distributions for 
$\cos\theta_\mu \in [0.9,1.0]$ (left panel) and 
$\cos\theta_\mu \in [0.4,0.5]$ (right panel)
in the three-flavor case,
and in scenarios with sterile neutrinos of two candidate $\dmssn$ values,
corresponding to the N-N-1 and N-N-2 configurations.
We take $|U_{e4}|^2=0.025$, $|U_{\mu 4}|^2=0.05$, and $|U_{\tau 4}|^2=0$.
The information on hadron energy is not used.
Note that the event distributions in the three-flavour scenario, and in 
the small-$\Delta m^2$ scenario are very close, and their difference
cannot be discerned in the figure.
}
\label{fig:events-with-energy}
\end{figure}
\begin{figure}[]
\centering
\includegraphics[width=0.49\textwidth]{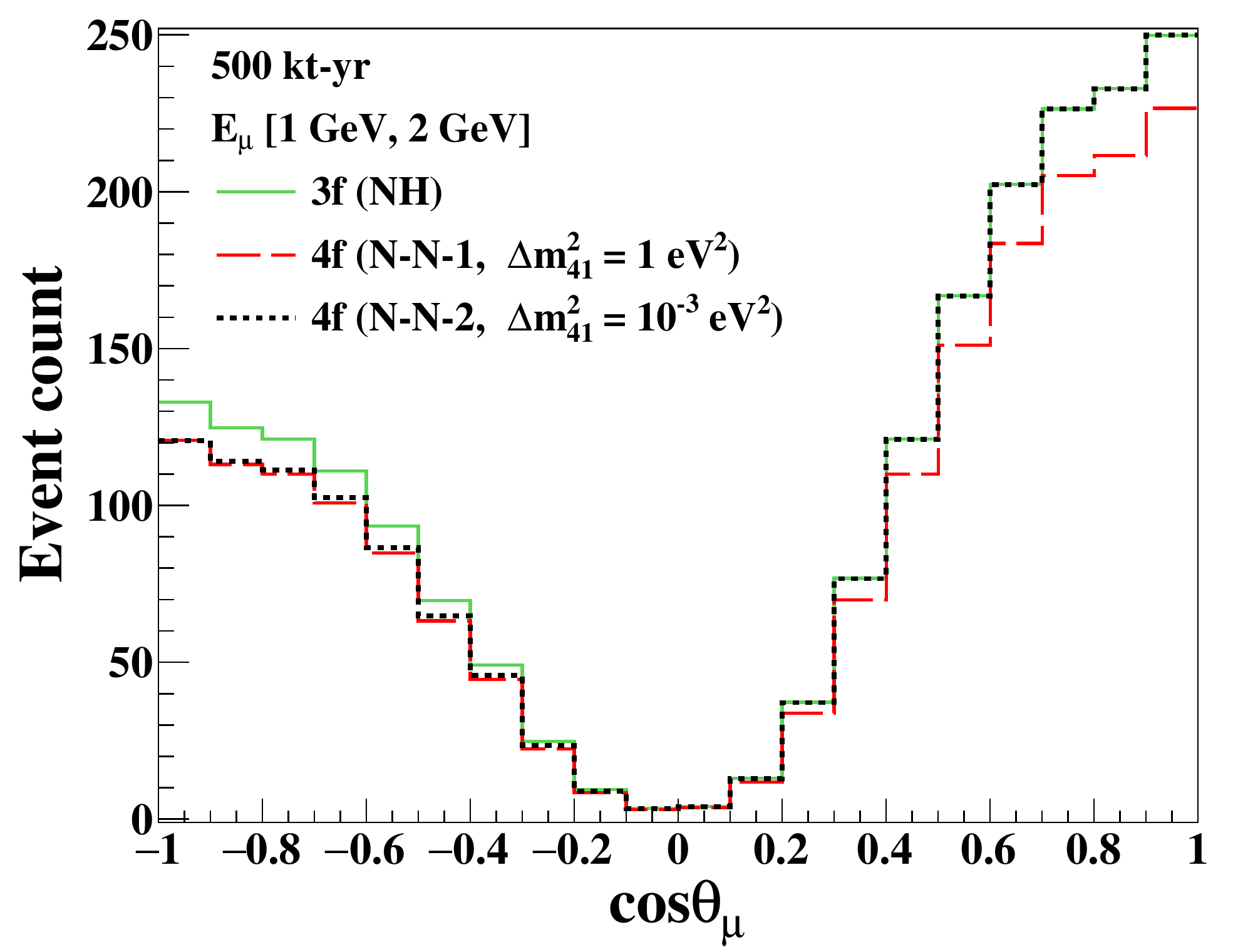}
\includegraphics[width=0.49\textwidth]{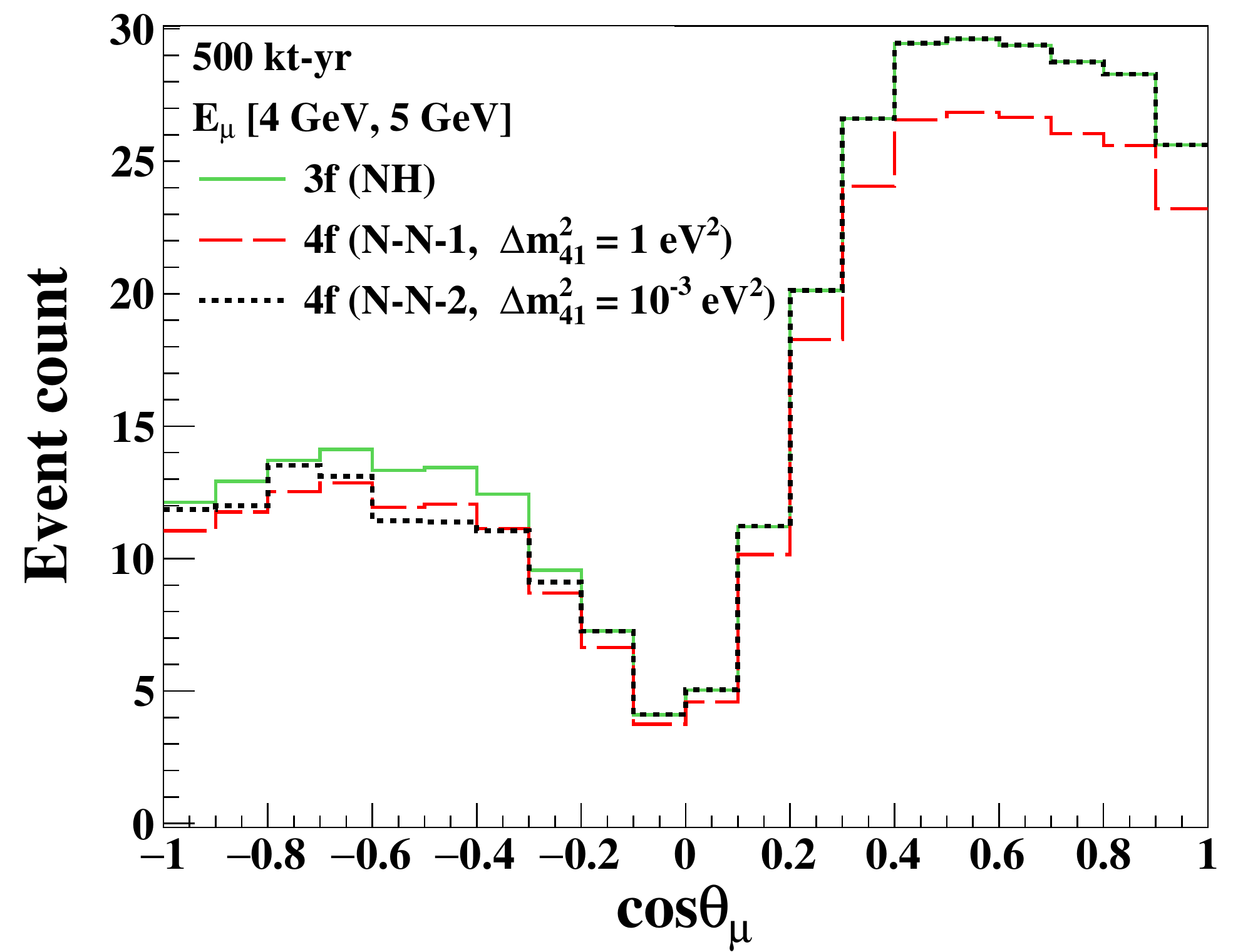}
\mycaption{$\mu^-$ event distributions for $E_\mu \in [1,2]$ GeV (left panel)
and $E_\mu \in [4,5]$ GeV (right panel) 
in the three-flavor case,
and in scenarios with sterile neutrinos of two candidate $\dmssn$ values,
corresponding to the N-N-1 and N-N-2 configurations.
We take $|U_{e4}|^2=0.025$, $|U_{\mu 4}|^2=0.05$, and $|U_{\tau 4}|^2=0$.
The information on hadron energy is not used.
}
\label{fig:events-with-costheta}
\end{figure}

Figure~\ref{fig:events-with-energy} shows the distribution of $\mu^-$ events 
as a function of $E_\mu$, the reconstructed muon energy in two sample 
$\cos\theta_\mu$-bins, where the efficiencies and resolutions of the detector 
are folded in. 
Clearly, the number of events should be smaller in the $4\nu$ mixing
scenario as compared to the $3\nu$ mixing scenario, due to the
nonzero $|U_{\mu 4}|$. The fractional
difference is almost uniform over energy for $\Delta m^2_{41}=1$ eV$^2$
(N-N-1 configuration).
The absolute difference in the number of events is the largest in the 
lowest energy bin and decreases at larger energies.
This indicates that the low-energy data will contribute more significantly
to the identification of sterile mixing.
For $\Delta m^2_{41}= 10^{-3}$ eV$^2$ (N-N-2 configuration), 
on the other hand, the difference in the number of events as a function of 
energy is small and the energy dependence is mild.
We have explicitly checked that the features for $\dm_{41}=10^{-5}$ eV$^2$ 
(N-N-3 configuration) are the same as those for N-N-2 configuration.

Figure~\ref{fig:events-with-costheta} shows the event distribution 
as a function of reconstructed $\cos\theta_\mu$ in two sample energy bins.
For $\Delta m^2_{41}=1$ eV$^2$ (N-N-1 configuration), 
the fractional loss in the number of
events is observed to be almost uniform throughout the $\cos\theta_\mu$ range. 
On the other hand, $\Delta m^2_{41}= 10^{-3}$ eV$^2$ (N-N-2 configuration)
yields almost no loss of downward-going events ($\cos\theta_\mu \gtrsim 0$)
as compared to the three-flavour scenario, while for the upward-going
events ($\cos\theta_\mu \lesssim 0$), a clear loss of events is observed. 
This is due to the fact that at such low $\Delta m^2_{41}$ values, the 
downward-going neutrinos do not have enough time to oscillate to the sterile 
flavor.
We have explicitly checked that the features for $\dm_{41}=10^{-5}$ eV$^2$ 
(N-N-3 configuration) are the same as those for N-N-2 configuration.
This non-trivial dependence of the event spectrum on the direction of 
the muon is instrumental in providing sensitivity to the value of 
$|U_{\mu 4}|$ at low $\Delta m^2_{41}$.

\subsection{Useful regions in the $E_\mu$--$\cos\theta_\mu$ plane}
\label{chisq-analysis}

We constrain the active-sterile neutrino oscillation by using
\begin{equation}
\Delta \chi^2 \equiv \chi^2 \mbox{(4f)} - \chi^2 \mbox{(3f)} \; .
\end{equation} 
      Here, we define the $\chi^2$ for the four-flavor
      as well as three-flavor scenarios as follows.

First we compute the $\chi^2_{-}$ for $\mu^{-}$ events. 
In the ``2D'' analysis, where the information on only $E_\mu$ and
$\theta_\mu$ is used, we define the $\chi^2$
as~\cite{Huber:2002mx,Fogli:2002au,GonzalezGarcia:2004wg}:
\begin{equation}
\chi^{2}_{- {\rm (2D)}}
        ={\min_{\xi_l}} \sum_{j=1}^{N_{E_{\mu}}}
\sum_{k=1}^{N_{\cos \theta_{\mu}}} \left[ 2(N_{jk}^{\rm theory} -  N_{jk}^{\rm data})
        - 2 N_{jk}^{\rm data} \: \ln \left( \frac{N_{jk}^{\rm theory}}
{N_{jk}^{\rm data}} \right) \right]
          + \sum_{l=1}^{5} \xi_{l}^{2}\,,
\label{chisq-2D}
\end{equation}
where
\begin{equation}
N^{\rm theory}_{jk} = N^{0}_{jk}\bigg(1 + \sum_{l=1}^{5} \pi_{jk}^{l} \xi_{l}\bigg)\,.
\label{n-theory-definition-2D}
\end{equation}
In Eq.~(\ref{chisq-2D}), $N_{jk}^{\rm theory}$ and $N_{jk}^{\rm data}$ 
denote the expected and observed number of $\mu^-$ events, respectively,
in a given ($E_{\mu}$, $\cos \theta_{\mu}$) bin~\cite{Ghosh:2012px}.
Here $N^{0}_{jk}$ represents the number 
of events without systematic errors; we have taken $N_{E_{\mu}}$ = 10 and 
$N_{\cos\theta_{\mu}}$ = 20 as mentioned earlier.
The quantities $\xi_{l}$ indicate the ``pulls'' due to the systematic uncertainties.
Following~\cite{Ghosh:2012px}, we have included five systematic
errors in our analysis: a) flux normalization error (20\%),
b) cross section error (10\%), c) tilt error (5\%), d) zenith angle
error (5\%), and e) an overall systematics (5\%).

While using the additional information on hadron energy 
(the so-called ``3D'' analysis in~\cite{Devi:2014yaa}), 
the Poissonian $\chi^2_{-}$ for $\mu^{-}$ events takes
the form:
\begin{equation}
\chi^{2}_{-} {\rm (3D)}
        ={\min_{\xi_l}} \sum_{i=1}^{N_{E'_{\rm had}}} \sum_{j=1}^{N_{E_{\mu}}}
\sum_{k=1}^{N_{\cos \theta_{\mu}}} \left[ 2(N_{ijk}^{\rm theory} -  N_{ijk}^{\rm data})
        - 2 N_{ijk}^{\rm data} \: \ln \left( \frac{N_{ijk}^{\rm theory}}
{N_{ijk}^{\rm data}} \right) \right]
          + \sum_{l=1}^{5} \xi_{l}^{2}\,,
\label{chisq-3D}
\end{equation}
where
\begin{equation}
N^{\rm theory}_{ijk} = N^{0}_{ijk}\bigg(1 + \sum_{l=1}^{5} \pi_{ijk}^{l} \xi_{l}\bigg)\,.
\label{n-theory-definition-3D}
\end{equation}
In Eq.~(\ref{chisq-3D}), $N_{ijk}^{\rm theory}$ and $N_{ijk}^{\rm data}$ 
indicate the expected and observed number of $\mu^-$ events, respectively, 
in a given ($E_{\mu}$, $\cos \theta_{\mu}$, $E'_{\rm had}$) bin.
$N^{0}_{ijk}$ stands for the number of events without systematic errors. 
Since we consider four $E'_{\rm had}$ bins, $N_{E'_{\rm had}}$ = 4.

For both the ``2D'' and ``3D'' analyses, the $\chi^2_{+}$ 
for $\mu^{+}$ events is computed following the same procedure described 
above. We add the individual contributions from $\mu^-$ and $\mu^+$ 
events to obtain the total $\chi^2$ in 3$\nu$ and 4$\nu$ schemes.

\begin{figure}
\centering
\includegraphics[width=0.49\textwidth]{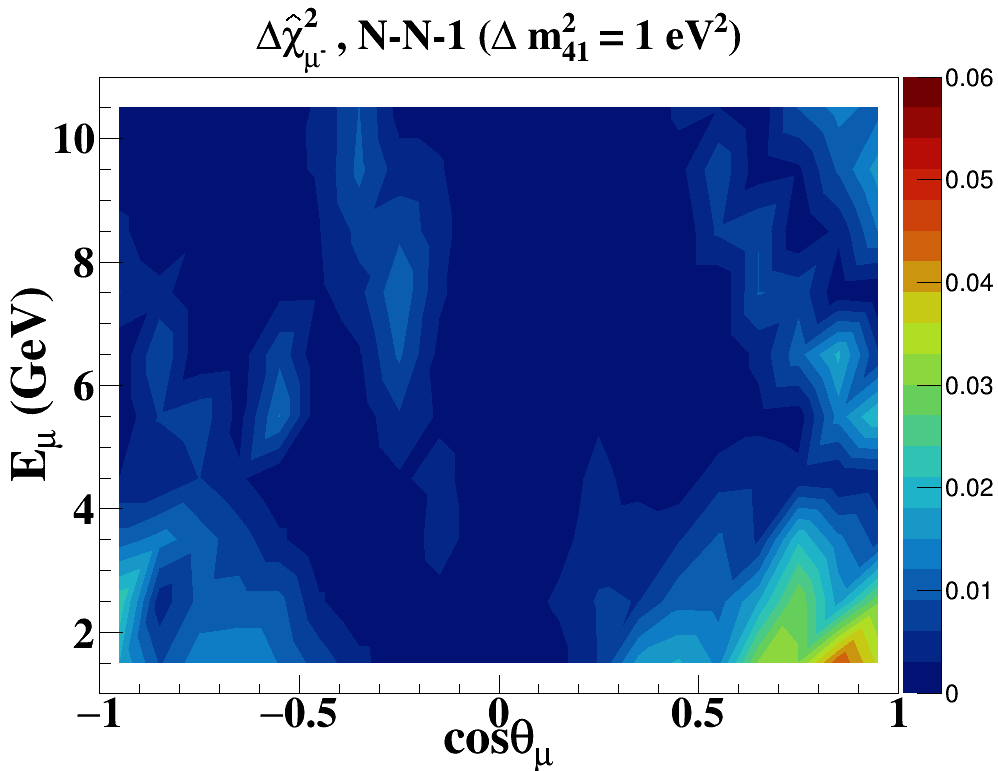}
\includegraphics[width=0.49\textwidth]{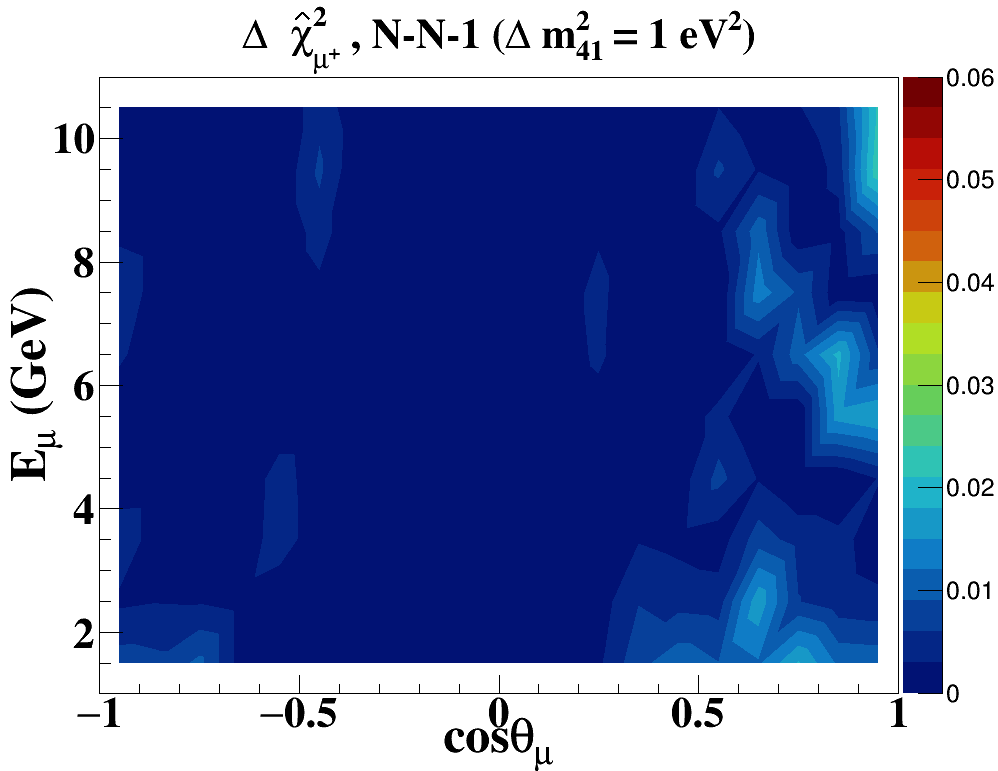} \\
\includegraphics[width=0.49\textwidth]{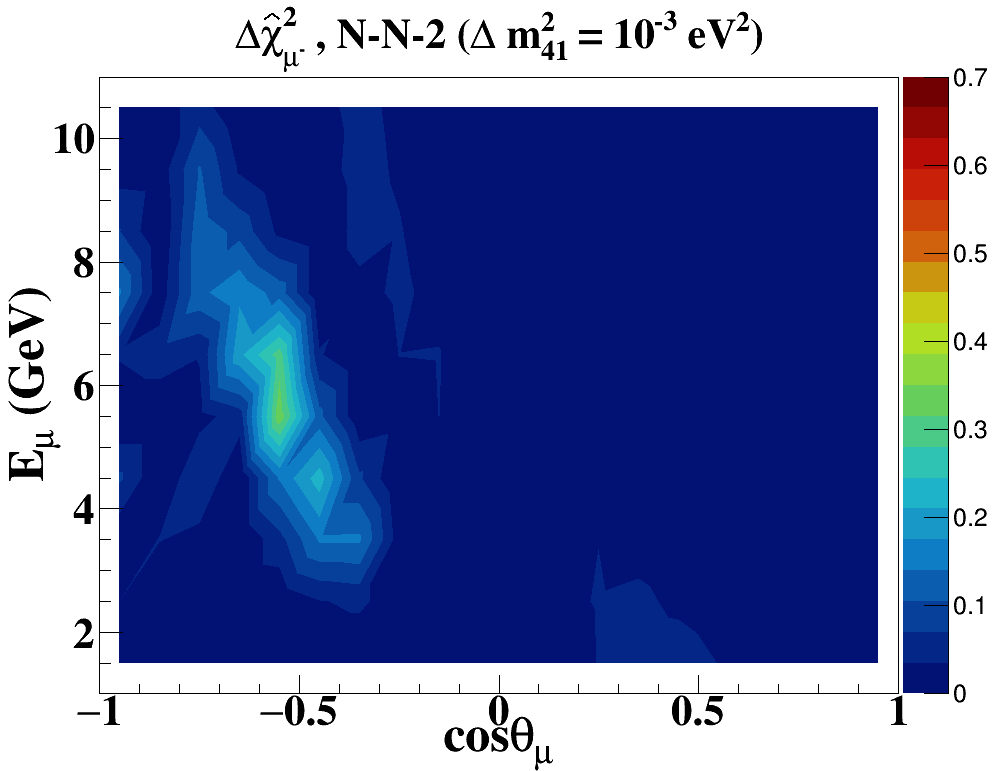}
\includegraphics[width=0.49\textwidth]{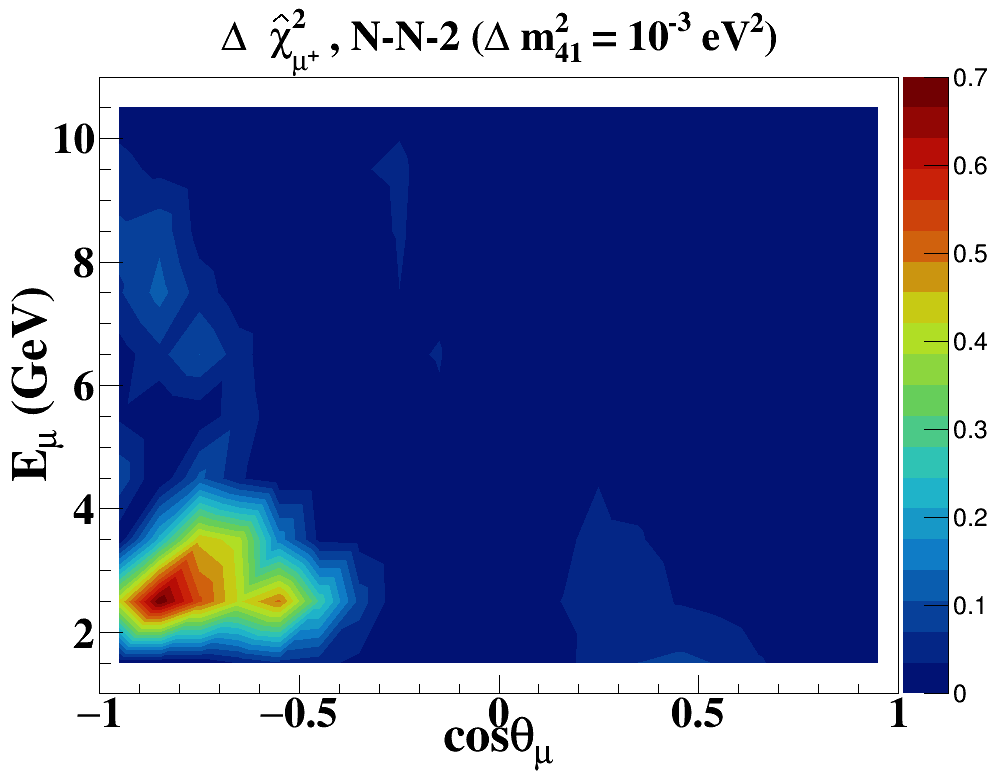}  
\mycaption{The distributions for $\Delta \widehat{\chi}^2$, i.e.
$\Delta\chi^2$ per energy-$\cos\theta$ interval,
for $\mu^-$ (left panel) and $\mu^+$ (right panel) events generated
using $|U_{e4}|^2=0.025$, $|U_{\mu 4}|^2=0.05$, and $|U_{\tau 4}|^2=0$.
The top panels correspond to $\Delta m^2_{41} = 1$ eV$^2$ (N-N-1 configuration),
while the bottom panels correspond to $\Delta m^2_{41} = 10^{-3}$ eV$^2$ 
(N-N-2 configuration).
Only the information on muon energy and direction is used.
Note that the scales in the top and bottom panels are different.}
\label{fig:chisq-distribution}
\end{figure}

In our analysis, the values of the six oscillation parameters of 
active neutrinos are taken to be fixed, in the simulated data as well 
as in the fit (see Table~\ref{tab:bench}). 
In order to judge the impact of $\theta_{23}$ and $\Delta m^2_{31}$,
the two parameters that are expected to influence the atmospheric neutrino
measurements strongly, we have also performed the analysis by marginalizing 
over these two parameters in their current $3\sigma$ allowed ranges.
We observe that, due to the marginalization over $\theta_{23}$ and 
$\Delta m^2_{31}$, the sensitivity of ICAL to sterile neutrino 
decreases marginally at low $\Delta m^2_{41}$, though it remains unchanged at 
$\Delta m^2_{41} \gtrsim 1$ eV$^2$. Even at lower $\Delta m^2_{41}$ values,
with the present relative $1\sigma$ precision of $\sim$ 5\%
on $\sin^2 2\theta_{13}$~\cite{Esteban:2016qun,Capozzi:2017ipn,deSalas:2017kay},
the impact of these marginalizations is very small and 
the results presented in this paper stay valid.
The other three parameters that appear in the
oscillation probabilities, viz. $\Delta m^2_{21}$, $\theta_{12}$, and
$\theta_{13}$, are known to a very good precision, and anyway appear in
the atmospheric neutrino oscillation probabilities as sub-leading terms
\cite{Akhmedov:2004ny}. Also, the impact of $\delta_{\rm CP}$ in the 
ICAL experiment is very mild~\cite{Kumar:2017sdq}.

Before we present our final sensitivity results on active sterile oscillation
parameters, we try to locate the regions in ($E_\mu$--$\cos\theta_\mu$) plane
which contribute significantly towards $\Delta \chi^2$.
In Fig.~\ref{fig:chisq-distribution},
we show the distribution of $\Delta \widehat{\chi}^2$, i.e. 
$\Delta\chi^2$ per energy-$\cos\theta$ interval (as in \cite{Devi:2014yaa}), 
in the reconstructed ($E_\mu$--$\cos\theta_\mu$) plane. 
Note that for this particular figure, we have not included the effects 
of the five systematic uncertainties mentioned above. 
The following insights are obtained from the figure.

\begin{itemize}

\item For $\Delta m^2_{41} = 1$ eV$^2$ (N-N-1 configuration), 
we expect there to be a overall suppression, with  
significant contributions coming from low energy downward going events. 
  This is because at such high values of $\Delta m^2_{41}$, active-sterile
oscillations develop even from downward neutrinos that travel $\sim$ 10 km
in the atmosphere, while the active $\nu$ oscillations would not have
developed. This feature would be prominent in the $\mu^{-}$ events,
but not so much in the $\mu^+$ events due to lack of statistics
(See Table~\ref{tab:CC-events1}).
Overall, we expect there to be an averaged overall suppression of events. 

\item For $\Delta m^2_{41} = 10^{-3}$ eV$^2$ (N-N-2 configuration), 
the information in 
  $\Delta\chi^2$ is concentrated in the low--energy upward-going events.
  This is because at low $\Delta m^2_{41}$, it takes $\sim$ 1000 km propagation 
  for the oscillations to develop. Also it may be noticed that the
  $\Delta \widehat{\chi}^2$ in this region is more significant for 
  $\mu^{+}$ events. This follows from the effects of the MSW resonance of
  active-sterile mixing in matter, which appears in the antineutrino channel
  for the N-N-2 configuration.

\item The events around $\cos\theta_\mu~\sim~0$ do not have much information,
  since the efficiency of the detector is less for horizontal
  events~\cite{Chatterjee:2014vta}.

\end{itemize}
 
In the next section, we discuss the constraints on the active-sterile mixing 
parameters in the $\dm_{41}$--$|U_{\mu 4}|^2$ space, and gain insights into the
dependence of these constraints on other mixing parameters like $|U_{e4}|$
and the mass ordering configuration.

\section{Constraining active-sterile mixing}
\label{sec:constraints}

In this section, we present the results on the reach of ICAL for
excluding the mixing parameter $|U_{\mu 4}|^2$, as a function of 
$\Delta m^2_{41}$. We generate the data in the absence of sterile 
neutrino, and try to fit it with the hypothesis of the presence of
a sterile neutrino, with $\Delta m^2_{41}$ ranging from $10^2$ eV$^2$
all the way down to $10^{-5}$ eV$^2$.
This range covers, and goes well beyond, the $\dm_{41}$ range relevant
for the LSND anomaly on the higher side and the solar $\dm_{21}$
on the lower side. We use an exposure of
500 kt-yr, and present the results in terms of 
90\% C.L. (2 d.o.f.) exclusion contours
in the $\Delta m^2_{41}$--$|U_{\mu 4}|^2$ plane. 
To generate the prospective data, we take the values of the 
active neutrino mixing parameters from Table~\ref{tab:bench} and
we keep them fixed in the fit. We have checked that, when the accuracy 
in the measurement of active mixing
parameters expected after 10 years is taken into account (using priors),
the variation in the values of these parameters does not affect the
results to any significant extent. We take $U_{\tau 4}$ as well as all the 
phases in the active-sterile mixing to be zero throughout the analysis, 
both in the simulation and in the fit.

In Sec.~\ref{excl-1}, to begin with we restrict ourselves to $|U_{e4}|$ = 0,
and the N-N ordering scheme ($\Delta m^2_{31} > 0$ and $\Delta m^2_{41} > 0$) 
for the sake of clarity.
In Sec.~\ref{exclue4nonzero}, we study the dependence of
the exclusion contours on nonzero $|U_{e4}|$
and find that the bounds in the ($\dm_{41}$,$|U_{\mu 4}|^2)$ parameter space
are the most conservative with $|U_{e4}|=0$.
We also check the dependence of the results on the mass ordering schemes
by computing the exclusion contours for the remaining ordering
schemes (N-I, I-N, and I-I), and find it to be very mild.
We therefore focus on the N-N
ordering scheme and keep $|U_{e4}|$ = 0, 
in the data as well as in the fit, in the rest of Sec.~\ref{sec:constraints}.
In Sec.~\ref{excl-rate}, we study the relative importance of the total rate of 
events and the shape of their energy as well as angular distributions,
and in Sec.~\ref{sec:comparison} we compare the INO-ICAL sensitivity
in the $\Delta m^2_{41}$--$|U_{\mu 4}|^2$ plane with that of the
other ongoing experiments.

\subsection{Exclusion contours in the 
$\Delta m^2_{41}$--$|U_{\mu 4}|^2$ plane}
\label{excl-1}

\begin{figure}
\centering
\includegraphics[width=0.95\textwidth]{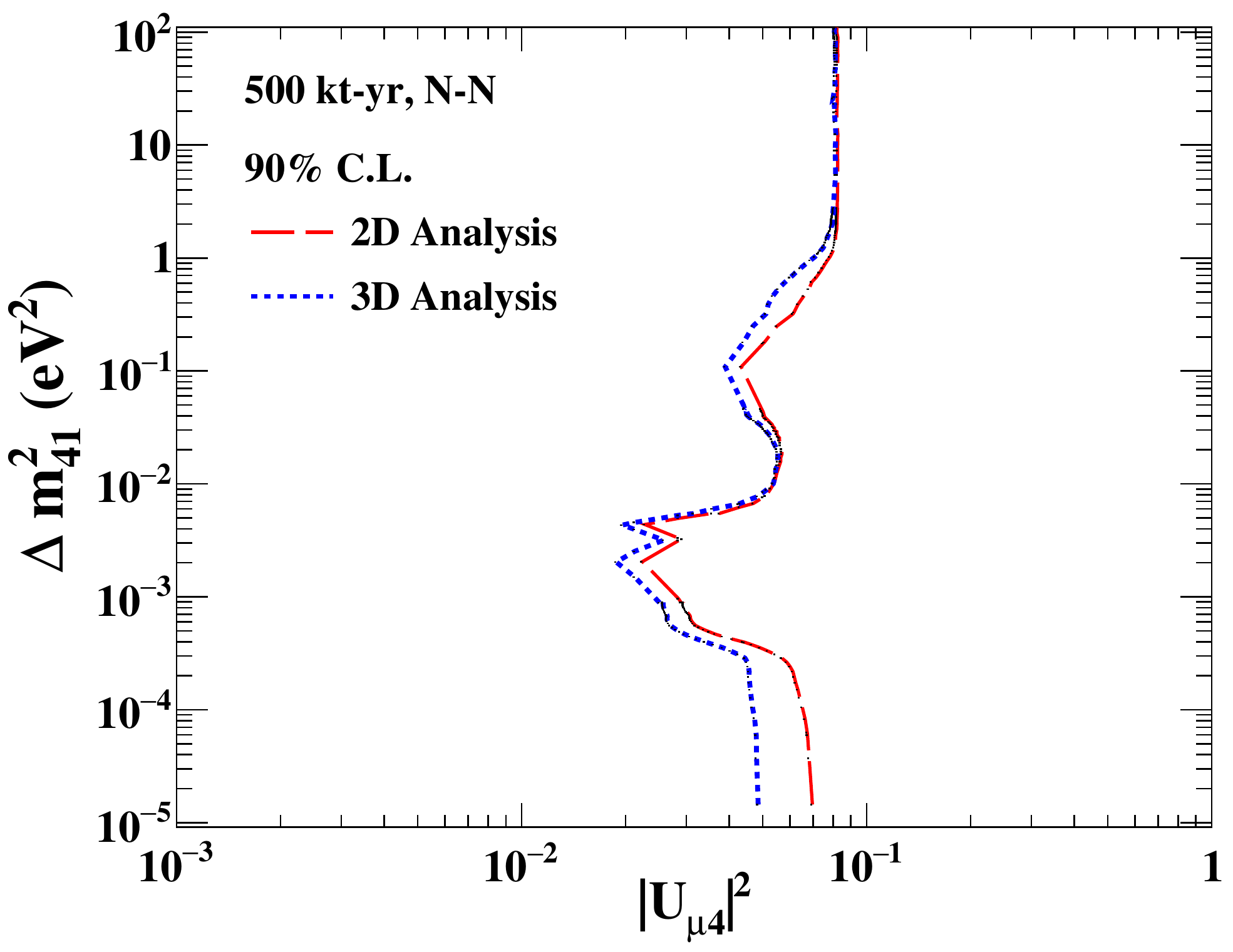}
\mycaption{The 90\% C.L. exclusion contours 
  in the $\Delta m^2_{41}$--$|U_{\mu 4}|^2$  plane, with the 2D analysis 
(information on muon momentum only), and 3D analysis (inclusion of
event-by-event information on hadron energy), for 500 kt-yr exposure. 
For illustration, we have used the N-N mass ordering scheme.
}
\label{fig:CC-exclusion-plots}
\end{figure}

Figure~\ref{fig:CC-exclusion-plots} shows the exclusion 
contours obtained using only the muon momentum information (2D analysis), 
and those obtained by adding the hadron energy information
(3D analysis), in the N-N mass ordering scheme.
The following features of the reach of ICAL,
expressed in terms of the exclusion contours, may be observed and
interpreted in terms of broad characteristics of neutrino oscillations.
 
\begin{itemize}

\item For $\Delta m^2_{41} \gtrsim 1$ eV$^2$, the ICAL can exclude
  active-sterile
mixing for $|U_{\mu 4}|^2 > 0.08$. The reach of ICAL in this region is
independent of the exact value of $\Delta m^2_{41}$.
  This is the 
high-$\Delta m^2_{41}$ region where neutrinos from all directions have
undergone many oscillations, such that only the averaged oscillation 
probability is observable, and there is no $\Delta m^2_{41}$-dependence.

\item  For $10^{-2}$ eV$^2 \lesssim \Delta m^2_{41} \lesssim 1$ eV$^2$, the 
downward-going neutrinos do not have enough time to oscillate, but the
upward-going neutrinos do. This non-trivial direction dependence improves the 
reach of ICAL to $|U_{\mu 4}|^2 > 0.05$ at $\Delta m^2_{41} \sim 10^{-1}$ eV$^2$.
  
\item
  Our results for $0.1$ eV$^2 \lesssim \dm_{41} \lesssim 10$ eV$^2$
    are consistent with those obtained in~\cite{Behera:2016kwr}.
    
\item In the region $10^{-4}$ eV$^2 \lesssim \Delta m^2_{41} 
\lesssim 10^{-2}$ eV$^2$, 
the oscillation frequencies $\widetilde{\Delta}_{41}$ and
  $\widetilde{\Delta}_{31}$ (where $~\widetilde{ }~$ denotes the quantity in the
  presence of matter) are of the same order of magnitude. 
As a result, interference effects between these frequencies leads to
a further improvement in the reach of ICAL in this
parameter range: around $\Delta m^2_{41} \sim 10^{-3}$ eV$^2$, the ICAL reach 
can be up to $|U_{\mu 4}|^2 > 0.03$.

\item When $\Delta m^2_{41} \lesssim 10^{-4}$ eV$^2$, the active-sterile 
  oscillations due to the frequency $\widetilde{\Delta}_{41}$
  will be suppressed.
  However in this case
  $\widetilde{\Delta}_{43}  \approx -\widetilde{\Delta}_{31}$,
  and the oscillations due to this frequency will continue to be present.
These oscillations will be independent of the value of $\Delta m^2_{41}$, 
and hence the reach of ICAL in this lowest $\Delta m^2_{41}$ range,
$|U_{\mu 4}|^2 \gtrsim 0.07$, would be virtually
independent of the exact $\Delta m^2_{41}$ value.

\item The addition of hadron information, as described in \cite{Devi:2014yaa},
results in a small improvement in the reach of ICAL for excluding $|U_{\mu 4}|$. 
The improvement is negligible for $\Delta m^2_{41} \gtrsim 1$ eV$^2$, the region
where the $\Delta\chi^2$ information is almost uniformly distributed in the
whole $E_\mu$- and $\cos\theta_\mu$-range, and hence additional details of
the hadron information do not make much difference. At lower $\Delta m^2_{41}$
values, though, the hadron information may contribute significantly. At the
lowest $\Delta m^2_{41}$ range, it increases the ICAL reach from
$|U_{\mu 4}|^2 \gtrsim 0.07$ to $|U_{\mu 4}|^2 \gtrsim 0.05$. 

\end{itemize}

\subsubsection{Dependence on $|U_{e4}|$ and the mass ordering scheme}
\label{exclue4nonzero}

\begin{figure}
\centering
\includegraphics[width=0.49\textwidth]{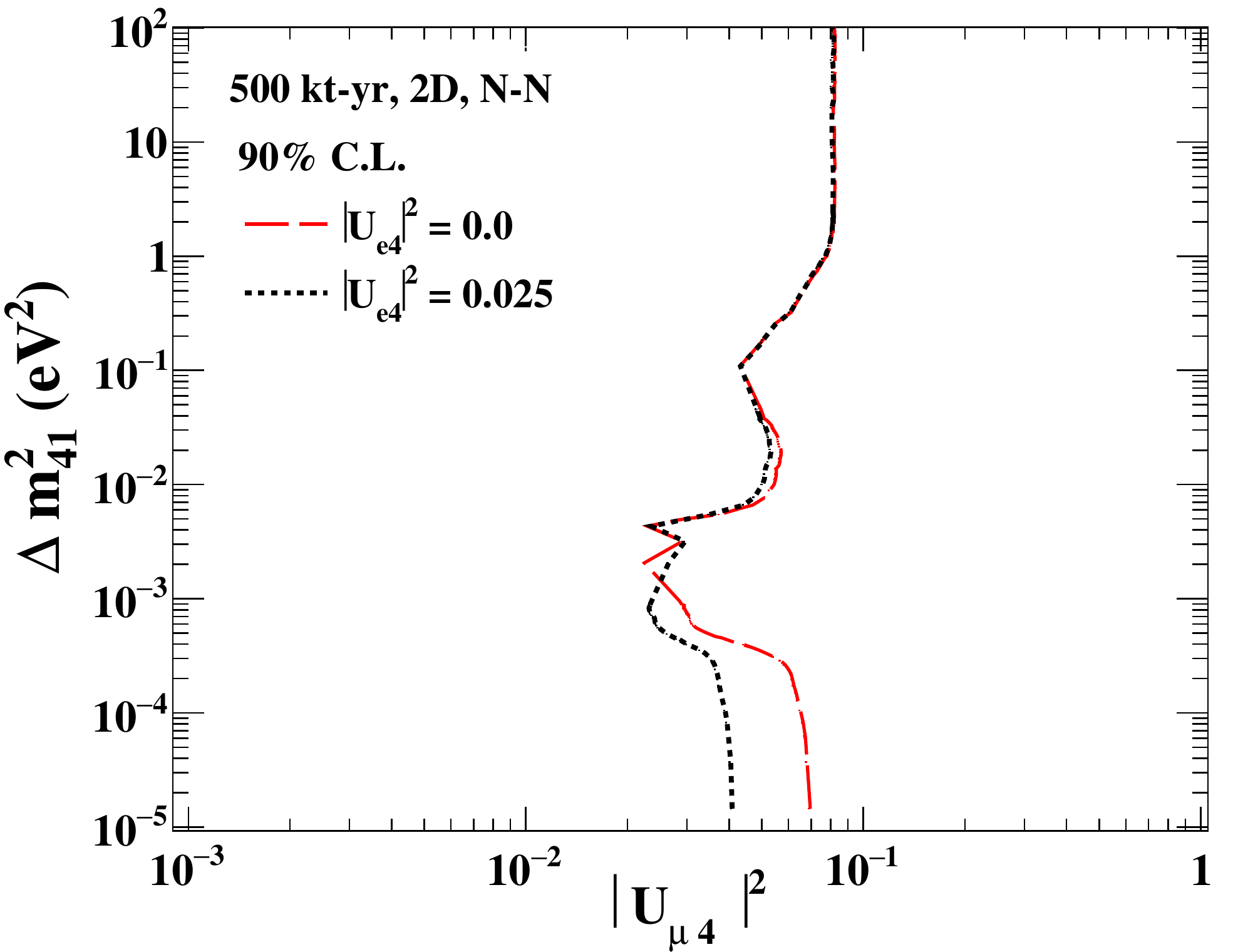}
\includegraphics[width=0.49\textwidth]{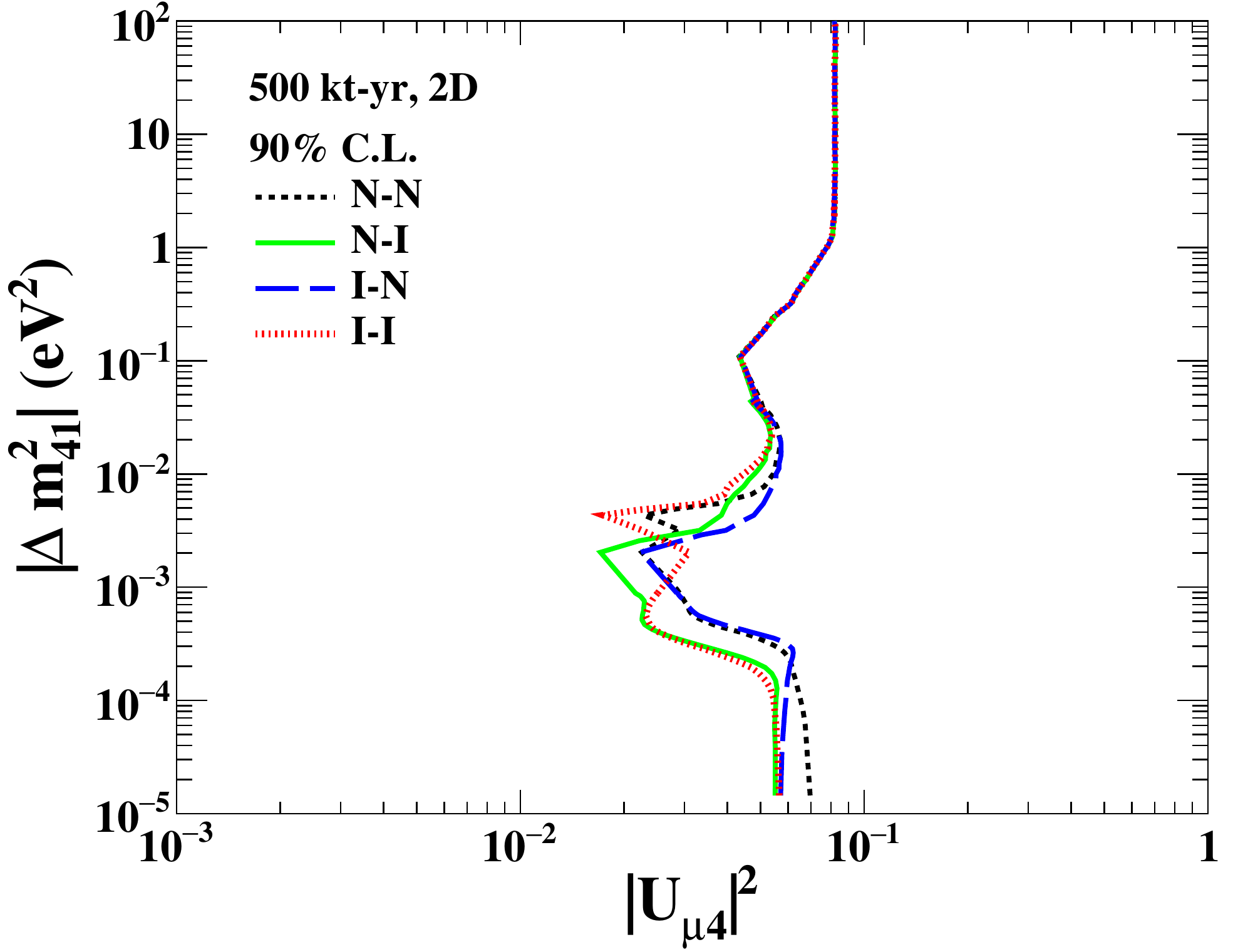}
\mycaption{Left panel: The effect of a nonzero value of $|U_{e4}|$ on the
  90\% exclusion contours, for 500 kt-yr exposure.
  Right panel: The effect of different mass ordering schemes
  on the sensitivity, for $|U_{e4}|=0$.
  The results have been presented for the N-N mass ordering scheme,
  and using only muon information.
}
\label{fig:CC-exclusion-plots1}
\end{figure}

While calculating the exclusion contours in Sec.~\ref{excl-1},
the value of $|U_{e4}|$ was taken to be zero.
A possible nonzero value of $|U_{e4}|$ could change the
results. This change may be discerned from 
the left panel of Fig.~\ref{fig:CC-exclusion-plots1}, 
where we show the change in the
exclusion contours when the value of $|U_{e4}|^2$ is taken to be 0.025.
The reach of ICAL clearly improves with a nonzero $|U_{e4}|$. The 
constraints given in the vanishing-$|U_{e4}|$ scenario are thus the 
most conservative, and in this paper, we continue to give
our constraints in this conservative limit.
In case the sterile neutrinos are discovered through the oscillations 
of $\bar\nu_e$ at reactor experiments, the measured value of $|U_{e4}|$
would help in improving the ICAL constraints.

The right panel of Figure~\ref{fig:CC-exclusion-plots1} shows the
dependence of the exclusion contours on the mass ordering scheme.
The figure suggests that the exclusion reach of ICAL does not depend on the
mass ordering scheme for $\Delta m^2_{41} \gtrsim 10^{-2}$ eV$^2$,
while it would depend mildly on the scheme for
$\Delta m^2_{41} \lesssim 10^{-2}$ eV$^2$.

\subsubsection{Contribution of rate, energy and direction
measurements}
\label{excl-rate}

\begin{figure}[t]
\centering
\includegraphics[width=0.49\textwidth]{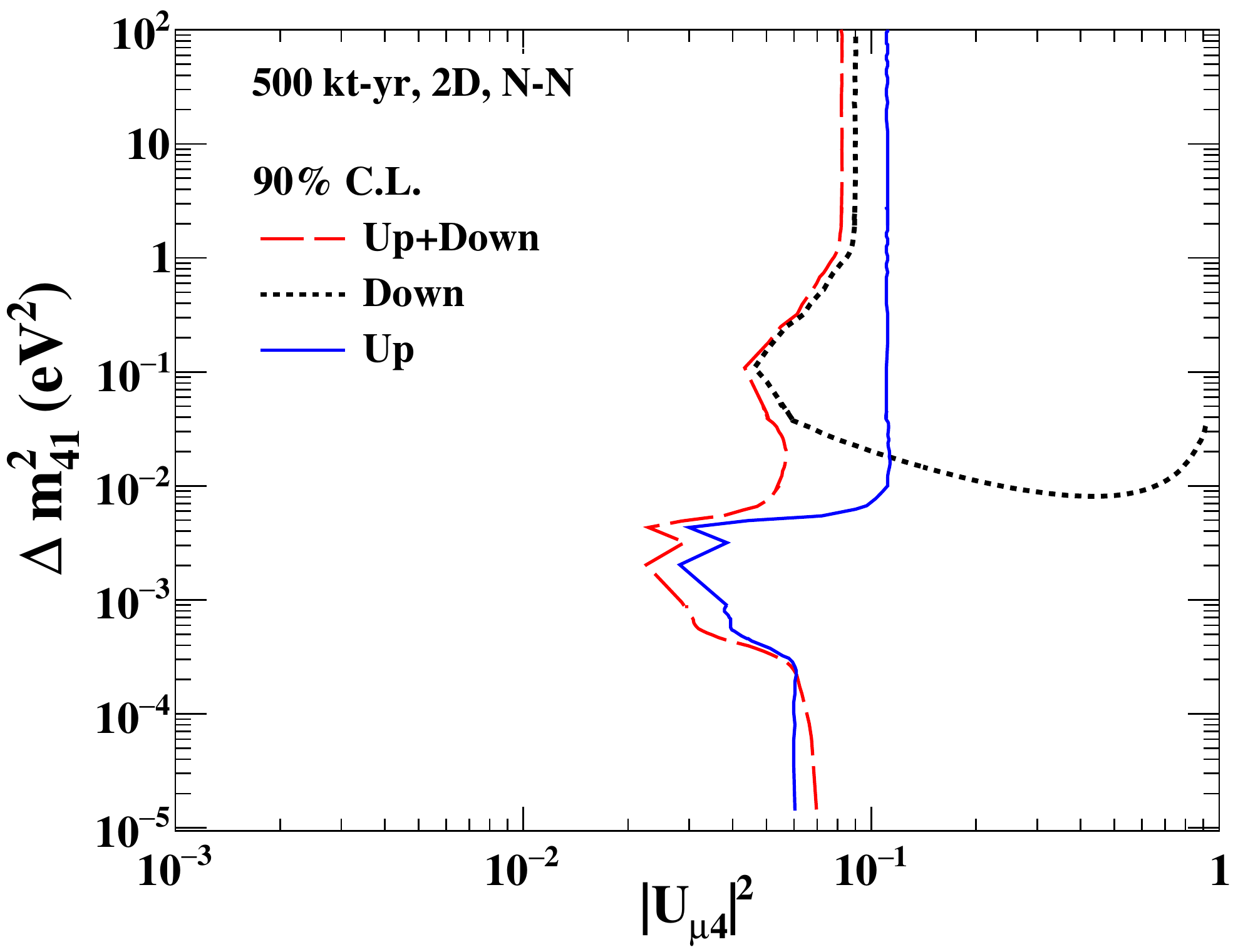}
\includegraphics[width=0.49\textwidth]{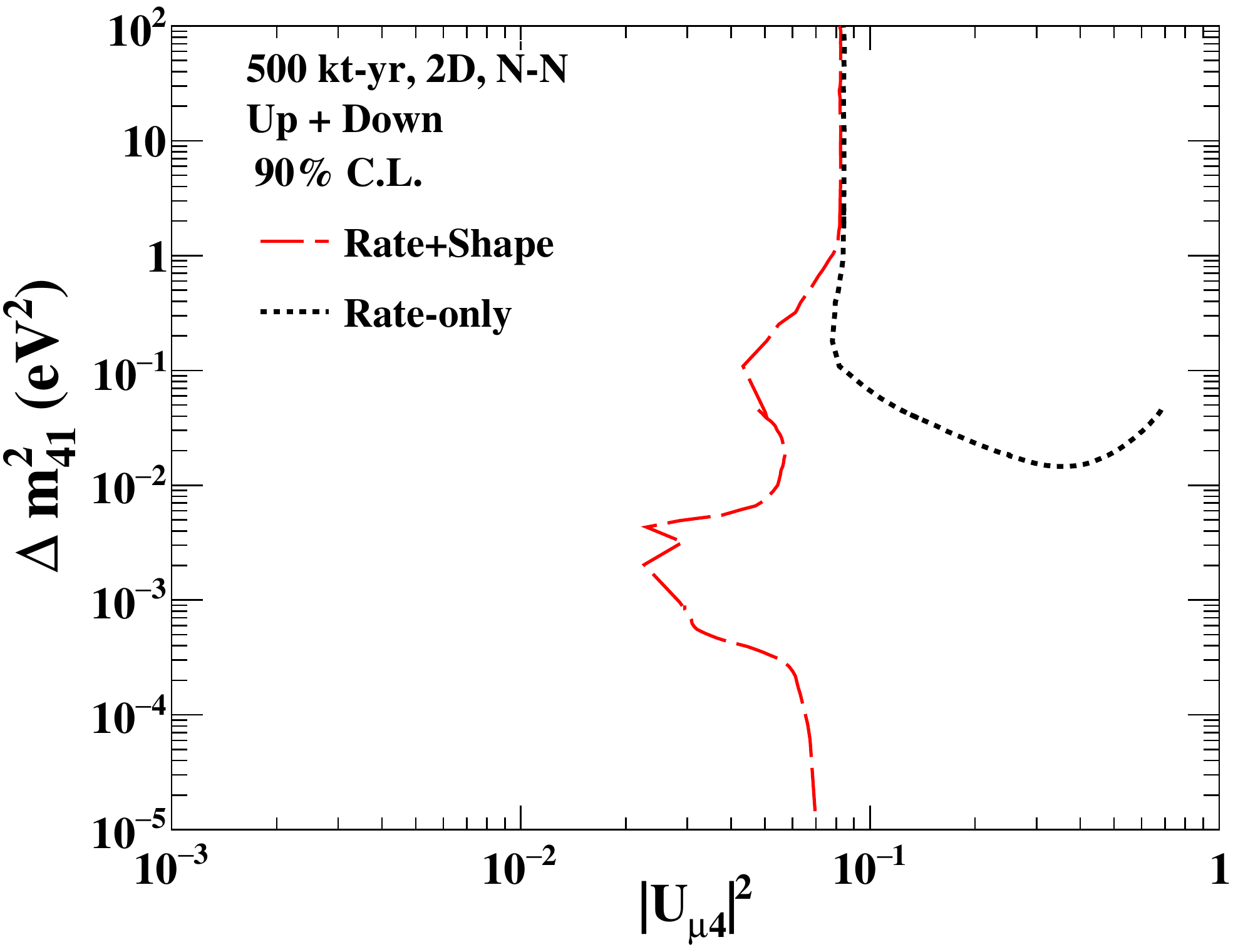}  
\mycaption{Left panel: 90\% C.L. exclusion regions in the 
$|U_{\mu 4}|^2$--$\Delta m^2_{41}$ plane with up-going events and 
down-going events separately, as well as their combined effect.
Right panel: 90\% C.L. exclusion regions in the 
$|U_{\mu 4}|^2$--$\Delta m^2_{41}$ plane using the information on both
the rate and shape of the muon spectrum, and using the information
only on the rate of muon events (integrated over the range [1--11] GeV
and over all the directions).
The results have been presented for the N-N mass ordering scheme,
and using only muon information.
}
\label{fig:rate-shape-comparison}
\end{figure}

The atmospheric data on charged-current muon neutrinos consist of the
event rates, as well as the energy and direction of the muon in these events. 
It is an instructive exercise, and an overall check of our analysis, to
compare the relative importance of these quantities, and interpret them
in terms of analytical approximations of oscillation probabilities.
In Fig.~\ref{fig:rate-shape-comparison}, we show the exclusion contours
if certain information were omitted, which would give us an idea of
how important that information is for the analysis. 

The left panel of Fig.~\ref{fig:rate-shape-comparison} indicates the 
relative importance of upward-going and downward-going events.
The right panel of the figure indicates the relative importance of the
event rate and the spectral shape (this includes the energy as well
as direction information for muons.)
The following observations may be made from these figures.

\begin{itemize}

\item For $\Delta m^2_{41} \gtrsim 1$ eV$^2$, the information
  about the active-sterile mixing is slightly more in the 
downward-going events than that in the upward-going
events. Since we know that in this parameter range the direction of
neutrinos does not matter, this difference should be simply due to the 
larger number of downward-going muon neutrinos (since about half the 
atmospheric $\nu_\mu$ are converted to $\nu_\tau$ on their way through 
the Earth). The reach of ICAL for $|U_{\mu 4}|^2$ values with the 
downward-going and upward-going data sets shown in the left panel 
is indeed observed to follow the approximate number of events in this range. 
This is also consistent with the observation from the right panel,
that in this $\Delta m^2_{41}$ range all the information is essentially
in the event rates, and almost no additional information comes from
the energies or directions of the muons.

\item  For $10^{-2}$ eV$^2 \lesssim \Delta m^2_{41} \lesssim 1$ eV$^2$, 
the left panel shows that the information from upward-going neutrinos is 
independent of $\Delta m^2_{41}$, which may be understood by observing that the
upward-going neutrinos in this $\Delta m^2_{41}$ range still undergo many 
oscillations inside the Earth, and only their averaged effect is observed 
at the detectors. On the other hand, the downward-going neutrinos undergo
a small number of oscillations ($\lesssim 1$) in the atmosphere of the Earth, 
and hence the exact value of $\Delta m^2_{41}$ is relevant. At the higher 
end of this $\Delta m^2_{41}$ range, most of the information about sterile
mixing is in the downward-going neutrinos, while at the lower end,
the information on sterile oscillations in down-going neutrinos is very small 
since these neutrinos do not get enough time to oscillate. 

\item For $10^{-2}$ eV$^2 \lesssim \Delta m^2_{41} \lesssim 1$ eV$^2$, 
the right hand panel shows results that support the above observation. 
At the lower end of this range, the loss of neutrinos due to oscillations
to the sterile species is small, as we have also observed in 
Sec.~\ref{sec:events}. So the information in only event rates is 
insignificant. On the other hand, most of the information is now in 
the neutrino direction (and hence the muon direction). 

\item At extremely low $\Delta m^2_{41}$ values, only the upward-going
neutrinos oscillate, and even their number of oscillations is small, so that
almost all the information is in the energy and direction distribution
of the events. 

\end{itemize}

The analysis in this section shows that the information on muon direction
is crucial for $\Delta m^2_{41} \lesssim 1$ eV$^2$. 
The accurate muon direction measurements at 
ICAL~\cite{Chatterjee:2014vta,Kumar:2017sdq} 
thus make it a suitable detector for probing low $\Delta m^2_{41}$ values.

\subsection{Comparison with other experiments}
\label{sec:comparison}

\begin{figure}[t]
\centering
\includegraphics[width=0.95\textwidth]{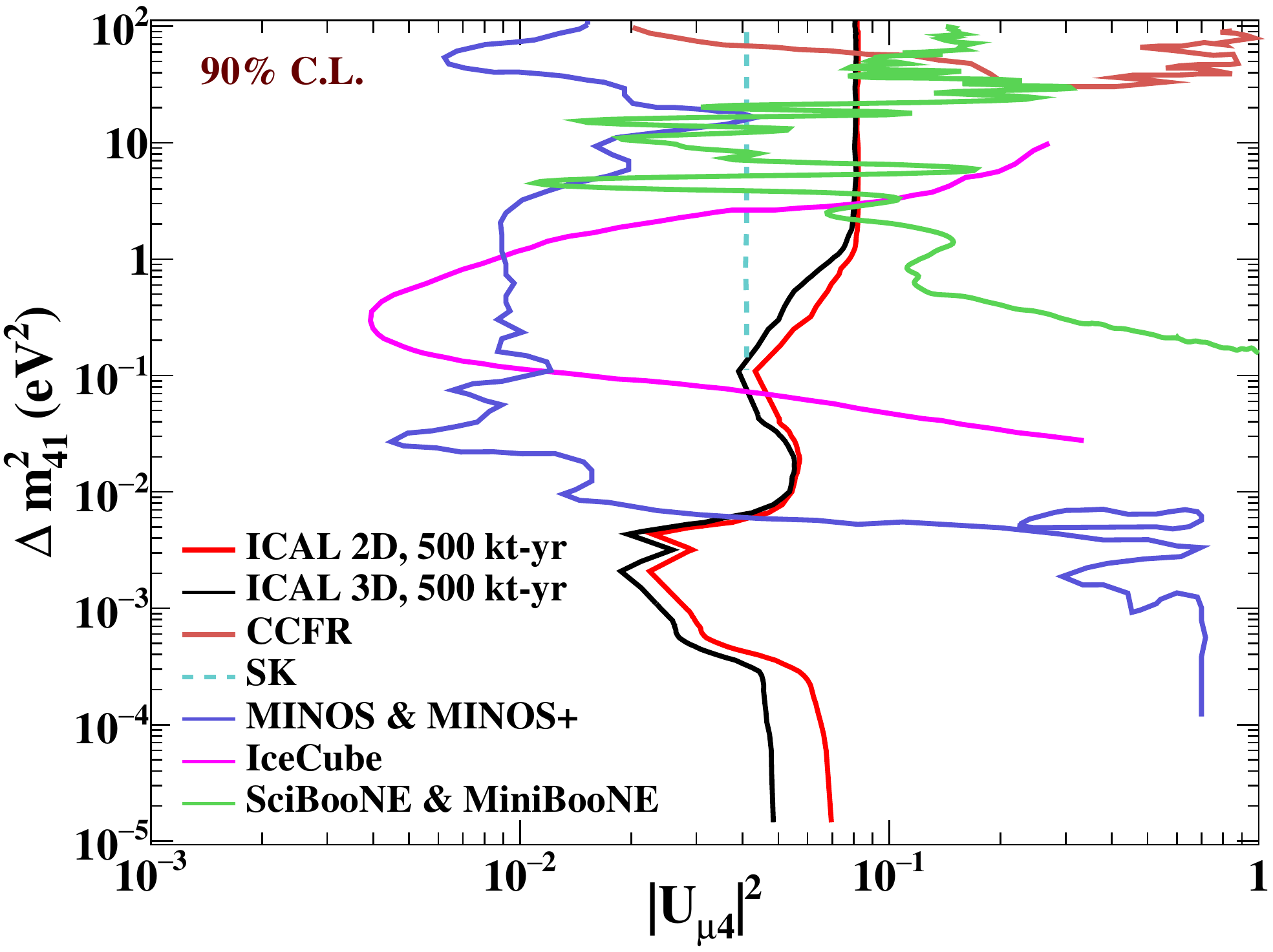}
\mycaption{
  Comparison of the 90\% exclusion contours for INO-ICAL in
  the N-N mass ordering scheme with an exposure of 500 kt-yr,
  with the current constraints from  other experiments:
  CCFR~\cite{Stockdale:1984cg},
  Super-Kamiokande (SK)~\cite{Abe:2014gda},
  IceCube~\cite{TheIceCube:2016oqi},
  MINOS and MINOS+~\cite{Adamson:2017uda},
  SciBooNE and MiniBooNE~\cite{Cheng:2012yy}.
  The Super-Kamiokande contour, shown with a dotted line,
  is the result of the analysis using
  a one-parameter fit, while all the other
  contours are with analyses using two-parameter fits.
  }
\label{fig:ical-comparison}
\end{figure}

Figure~\ref{fig:ical-comparison} shows the projected 90\% C.L. reach of 
ICAL for $|U_{\mu4}|^2$ with 500 kt-yr of data, and its comparison with
the current bounds from the atmospheric neutrino experiments
Super-Kamiokande~\cite{Abe:2014gda} and Icecube~\cite{TheIceCube:2016oqi},
the short-baseline experiments CCFR~\cite{Stockdale:1984cg},
MiniBooNE and SciBooNE~\cite{Cheng:2012yy}, as well as the long-baseline
experiments MINOS and MINOS+~\cite{Adamson:2017uda}. 

Super-Kamiokande (SK) is the experiment most similar to the INO-ICAL, 
in the sense that it is sensitive to the atmospheric neutrinos in a similar energy range.
In addition to the $\nu_\mu \to \nu_\mu$ and $\nu_e \to \nu_\mu$
oscillation channels that ICAL is sensitive to, SK can also detect the
charged-current interactions of $\nu_e$, and hence can analyze
the $\nu_e \to \nu_e$ and $\nu_\mu \to \nu_e$ channels.
(However, in the SK exclusion limits shown in the figure, the information
from $\nu_\mu \to \nu_\mu$ channel dominates.)
It also has a lower energy threshold than ICAL, so it has the advantage
of being able to detect the low energy events where the depletion
in the number of events is the most prominent for higher $\dm_{41}$
(for example, at $\dm_{41} \gtrsim 1$ eV$^2$,
see Fig.~\ref{fig:events-with-energy}).
On the other hand, it cannot distinguish neutrinos from antineutrinos,
and does not have as good muon direction resolution as ICAL,
which would play a crucial role in identifying sterile neutrinos
for lower $\dm_{41}$ values (see Fig.~\ref{fig:events-with-costheta}).

An important difference between the sterile neutrino analysis of
SK and the other experiments needs to be noted. Since the SK analysis
has been performed only for
$\dm_{41} > 0.1$ eV$^2$, where the oscillations from all directions
are averaged out, the data is not sensitive to the value of $\dm_{41}$.
The fit in the analysis is therefore carried out for only one
parameter ($|U_{\mu 4}|^2$), and therefore the 90\% exclusion contour
corresponds to $\Delta \chi^2 > 2.71$. On the other hand, for all the
other experiments including ICAL, the 90\% exclusion contours have
been evaluated with two-parameter analyses, and hence
correspond to $\Delta \chi^2 > 4.61$.
If the sensitivity analysis of ICAL were to be performed only in the
$\dm_{41} > 1$ eV$^2$ region with a single-parameter fit, the
90\% exclusion reach of ICAL would improve from
$|U_{\mu 4}|^2 > 0.08$ to $|U_{\mu 4}|^2 > 0.06$.

Some of the most stringent bounds on the sterile mixing parameters
are given by IceCube \cite{TheIceCube:2016oqi}. Indeed for
$\dm_{41} \approx 0.3$ eV$^2$, IceCube puts a bound of
$|U_{\mu 4}|^2 \lesssim 0.004$. However since IceCube has a lower
energy threshold of $\sim 100$ GeV, it is sensitive to only higher
values of $\dm_{41} \gtrsim 10^{-2}$ eV$^2$, as is evident
from Fig.~\ref{fig:ical-comparison}.

The fluxes at the fixed-baseline experiments are more well-determined
than the atmospheric neutrino fluxes at ICAL.
However ICAL has the advantage of a large range of $L/E$, and hence
sensitivity to a large range of $\Delta m^2_{41}$ values.
Indeed, this paper shows how qualitatively
different features of oscillation probabilities come into play for different
$\Delta m^2_{41}$ values, and how ICAL can address all these features so
as to be sensitive to a large range of $\Delta m^2_{41}$.
The CCFR experiment at Fermilab \cite{Stockdale:1984cg} had large neutrino
energies ($\gtrsim 100$ GeV) and a small baseline ($\sim 1$ km),
and could only probe $\Delta m^2_{41} \gtrsim 1$ eV$^2$.
The SciBOONE and MiniBOONE experiments
have nearly the same baselines as CCFR, but neutrino 
energies of $\sim$ GeV, so they could be sensitive 
to $\Delta m^2_{41} \gtrsim 0.1$ eV$^2$, 
but no lower~\cite{Mahn:2011ea}.
MINOS \cite{Adamson:2017uda}, having a longer baseline of $\sim 735$ km, 
is sensitive to even lower $\Delta m^2_{41}$ values, up to 
$\Delta m^2_{41} \gtrsim 10^{-4}$ eV$^2$.
At $\Delta m^2_{41} \lesssim 10^{-2}$ eV$^2$, ICAL would be 
one of the few experiments sensitive to active-sterile neutrino mixing.

\section{Exploring features of $\dm_{41}$}
\label{sec:discovery}

In the previous section, we derived the constraints in the
$\dm_{41}$--$|U_{\mu 4}|^2$ space in the scenario where ICAL observes
no signal for active-sterile oscillations. 
In this section, we assume that there is a sterile neutrino with
$10^{-5}$ eV$^2 \leq \Delta m^2_{41} \leq 10^2$ eV$^2$, 
with the mixing parameters $|U_{e4}|^2 = 0.025$, $|U_{\mu 4}|^2 = 0.05$
and $|U_{\tau 4}|^2 = 0$.
Note that while exploring the features of $\dm_{41}$ in this section,
we keep the values of the mixing parameters fixed while generating
the data in the $4\nu$ scenario.
  Even while fitting the data at test values of $\dm_{41}$, we take the
  active-sterile mixing parameters to be fixed and known.
  All the phases in the $4\nu$ mixing matrix are taken to be 
  zero in the data as well as in the fit.
  The values of the $3\nu$ mixing parameters have been kept fixed
  to the values shown in Table~\ref{tab:bench}, and 
  the sign of $\dm_{31}$ has been taken to be positive, while
generating the data and performing the fit.
In this situation, we attempt to address the following two questions
in the context of ICAL for the first time:
\begin{itemize}
\item With what precision can we measure a given value of $\dm_{41}$?
\item Can we identify the sign of $\dm_{41}$?
\end{itemize}
In this process, we will also try to quantify the contributions of two of the
unique features of ICAL, viz. charge-identification (CID) capability and
the hadron energy measurement on the event-by-event basis.

\subsection{Precision in the determination of $\Delta m^2_{41}$}
\label{sec:accuracy}

In order to illustrate the capability of ICAL to measure the
value of $\dm_{41}$ precisely, we generate data for an exposure
of 500 kt-yr at $\Delta m^2_{41}=10^{-3}$ eV$^2$,
which corresponds to the N-N-2 configuration,
The quantity
\begin{equation}
\Delta \chi^2_{\rm SP}[\Delta m^2_{41} \mbox{(test)}] \equiv  
\chi^2[\Delta m^2_{41} \mbox{(test)}] -  
\chi^2[\Delta m^2_{41}=10^{-3} \mbox{ eV}^2]
\label{eq:chisq-accuracy}
\end{equation}
is then calculated for test values of $\Delta m^2_{41}$ in the
whole range $10^{-5}$ eV$^2 < \dm_{41} < 0.1$ eV$^2$,
$-0.1$ eV$^2 < \dm_{41} < -10^{-5}$ eV$^2$,
which gives us
an idea of how well a particular value of $\dm_{41}$ may be excluded
from the data, if indeed there is a sterile neutrino.

\begin{figure}
\centering
\includegraphics[width=0.98\textwidth]{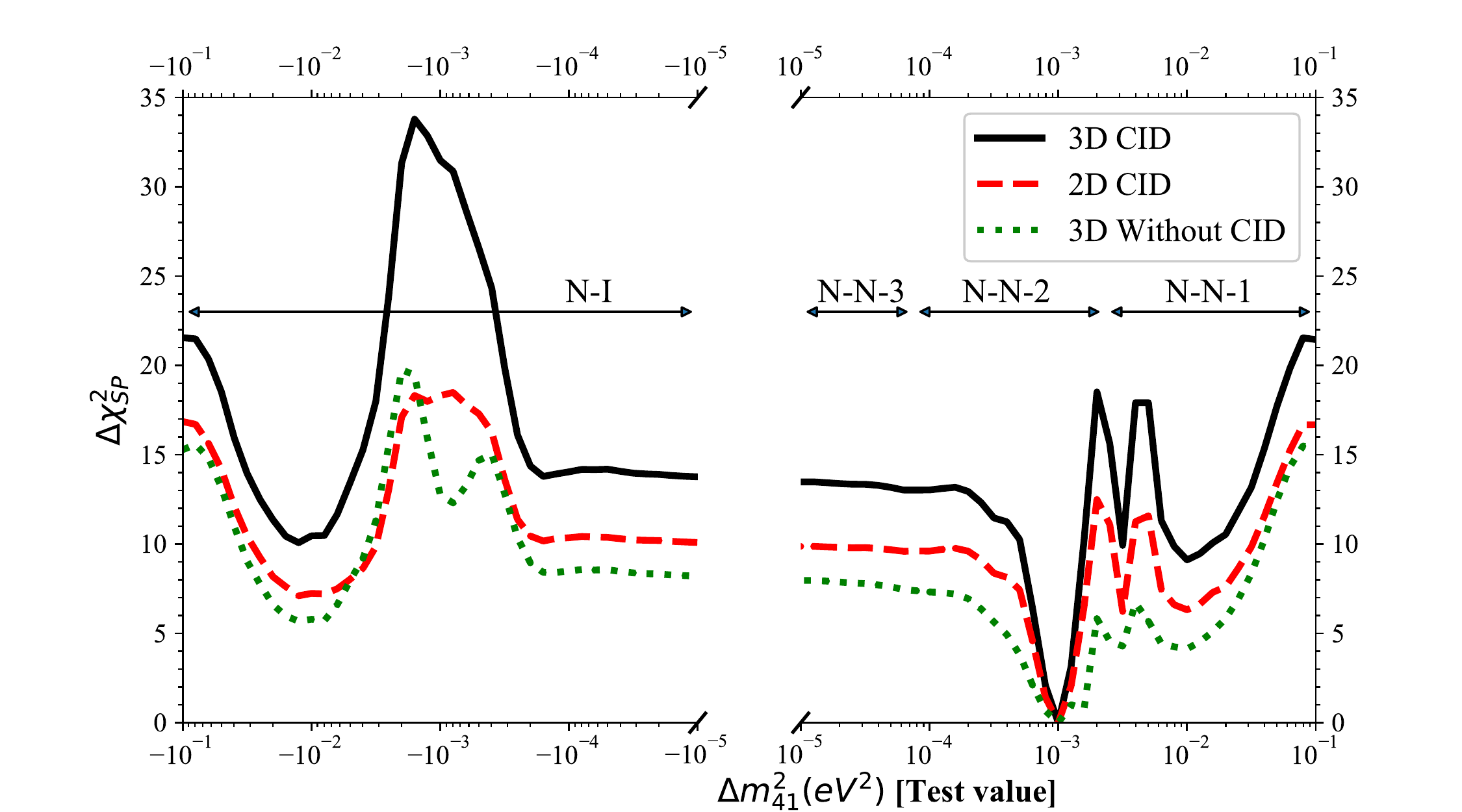}
\caption{The $\Delta\chi^2_{\rm SP}$ [see eq.~(\ref{eq:chisq-accuracy})]
at test values of $\Delta m^2_{41}$ with 500 kt-yr of ICAL data,
when the true value is $\Delta m^2_{41} = 10^{-3}$ eV$^2$ (corresponding 
to N-N-2). The mixing parameters are taken to be 
$|U_{e4}|^2 = 0.025$, $|U_{\mu 4}|^2 = 0.05$, and $|U_{\tau 4}|^2 = 0$.
The results have been presented with and without muon charge
identification, and with and without using the hadron energy
information.
}
\label{fig:dmsq-accuracy}
\end{figure}

The results have been shown in Fig.~\ref{fig:dmsq-accuracy}
for three different cases:
(i) 3D CID: information on muon charge identification (CID)
as well as hadron energy is used;
(ii) 2D CID: information on CID is used, but not on the hadron energy; and
(iii) 3D without CID: information on hadron energy is used, but not on CID. 
The muon momentum information is naturally used in all cases.
The comparison of these three cases will help us quantify the importance of
hadron energy calibration and muon charge identification,
two of the unique features of ICAL.
The following observations may be made from this figure.  

\begin{itemize}

\item The value of $\Delta \chi^2_{\rm SP}$ indeed shows a sharp dip near the
  true value of $\Delta m^2_{41}$. For the benchmark parameter values,
  it is expected that the other candidate mass ordering configurations,
  N-N-1, N-N-3 and N-I, could be ruled out at $\gtrsim 3\sigma$ in the 3D CID
  case.

\item Note that there is a narrow dip in $\Delta \chi^2_{\rm SP}$ near
  $\dm_{41} \approx (3$--$4) \times 10^{-3}$ eV$^2$. This is likely
  to be an effect of the interference between the oscillation frequencies 
  corresponding to $\dm_{41}$ and $\dm_{31}$.  

\item It is observed that the values of $\dm_{41} \approx -3 \times 10^{-3}$
  would be strongly disfavoured. The reason for this lies in the observation
  that in such a scenario, there would have been an active-sterile MSW
  resonance, that would have given rise to significant differences from
  the actual $\dm_{41} \approx +3 \times 10^{-3}$ scenario, which has no
  such resonance.

\item For $|\dm_{41}| \gtrsim 10^{-2}$ eV$^2$, the $\Delta \chi^2_{\rm SP}$ values
  are independent of the sign of $\dm_{41}$, since the active-sterile
  oscillations get averaged out at such a high magnitude of $\dm_{41}$,
  and the information about the sign of $\dm_{41}$ is lost.

\item In the absence of hadron energy information (2D CID),
  the value of $\Delta \chi^2_{\rm SP}$ 
  reduces by more than
  25\% at almost all test values of $\dm_{41}$. This illustrates the
  importance of using the hadron energy information, even though the
  measurement of hadron energy is not very accurate in ICAL.

\item Similarly, the absence of muon charge identification (3D without CID)
  would also reduce $\Delta \chi^2_{\rm SP}$  by more than 25\% at almost all test values
  of $\dm_{41}$. In particular, it would be difficult to rule out the N-N-3
  configuration, or the N-N-1 configuration with $\dm_{41} \lesssim 0.05$
  eV$^2$, to more than $3\sigma$ if the CID capability were absent.
  This illustrates the significant advantage ICAL would have due to its
  excellent muon charge identification.

\item The CID capability would also
  allow us to analyze the $\mu^-$ and $\mu^+$ events (equivalently,
  $\nu_\mu$ and $\bar\nu_\mu$) separately, and determine the $\Delta m^2_{41}$ 
  values with two independent data sets. This would also serve as a test
  of CPT violation in the $4\nu$ framework. 

\end{itemize}

As has been observed above, ICAL may be able to determine
$\dm_{41}$ to an accuracy that would enable us to identify the 
correct $m_i^2$ configuration from N-N-1, N-N-2, N-N-3, and N-I.
In particular, for $\dm_{31} > 0$ and $\dm_{41} \approx 10^{-3}$ eV$^2$,
it may be able to identify the sign of $\dm_{41}$.
In the next section, we explore this possibility further, focusing on
the feasibility of such an identification, with true $\dm_{41}$
taken over a wide range.

\subsection{Determination of the sign of $\Delta m^2_{41}$}
\label{sec:sign}

In the last section, we observed that ICAL is expected to be quite
sensitive to the mass ordering configurations in the four-neutrino
mass spectrum.
Earth matter effects would play a significant role in this identification,
since whether the active-sterile resonance takes place in the neutrino or
antineutrino channel depends crucially on the sign of $\dm_{41}$.
Matter effects thus magnify or suppress the effects due to
active-sterile mixing, and make the atmospheric data sensitive to
the mass ordering in the sterile sector. 

In this section, we shall focus on quantifying the sensitivity of ICAL to
the sterile sector mass ordering, i.e. the sign of $\dm_{41}$.
Fig.~\ref{fig:dmsq-accuracy} already shows that for $\dm_{41}=10^{-3}$ eV$^2$,
the wrong mass ordering would be ruled out to $\Delta\chi^2 \gtrsim 10$
with 500 kt-yr of exposure. We shall now examine how this result depends on
the actual value of $\dm_{41}$.

\begin{figure}[t]
\centering
\includegraphics[width=0.98\textwidth]{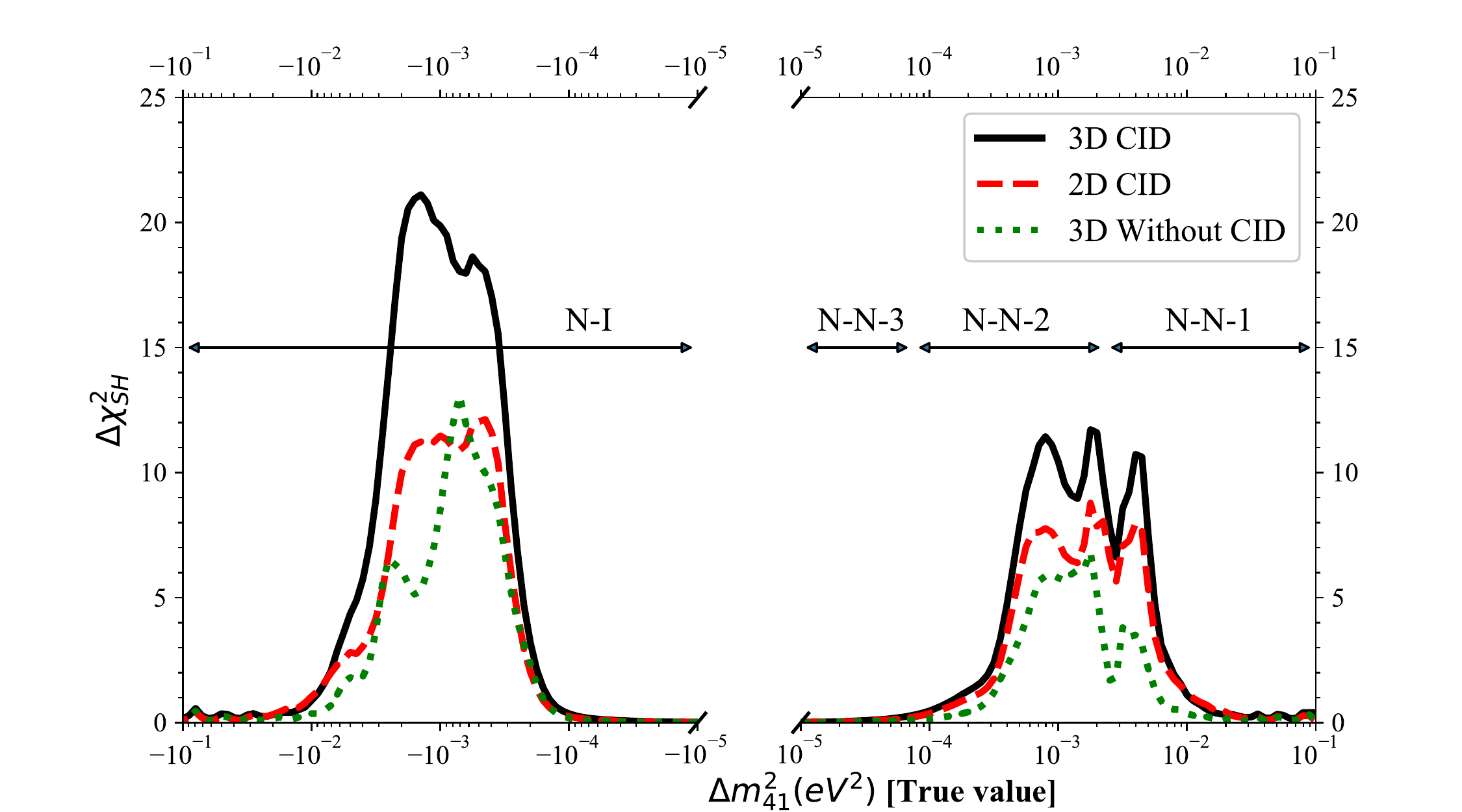}
\caption{The potential for identification of the sign of $\Delta m^2_{41}$
with 500 kt-yr of ICAL data, quantified by $\Delta\chi^2_{\rm SH}$ 
[see eq.~(\ref{eq:chisq-hierarchy})].
The mixing parameters are taken to be 
$|U_{e4}|^2 = 0.025$, $|U_{\mu 4}|^2 = 0.05$, and $|U_{\tau 4}|^2 = 0$.
The results have been presented with and without muon charge
identification, and with and without using the hadron energy
information.
}
\label{fig:sterile-hierarchy}
\end{figure}

We define the ICAL sensitivity to sterile neutrino hierarchy with the
quantity
\begin{equation}
\Delta \chi^2_{\rm SH} \equiv  
\chi^2_{\mbox{min}}[\Delta m^2_{41} \mbox{(wrong sign)}] -  
\chi^2[\Delta m^2_{41} \mbox{(true)}] \; , 
\label{eq:chisq-hierarchy}
\end{equation}
where $\chi^2_{\mbox{min}}[\Delta m^2_{41} \mbox{(wrong sign)}]$
denotes the minimum value of $\chi^2$ when the test value of
$\Delta m^2_{41}$ is varied over all values with the wrong sign.
Note that it is not enough to distinguish $\Delta m^2_{41}$
from $-\Delta m^2_{41}$, or some other effective value. We would
like to ensure that all possible values of $\Delta m^2_{41}$
with the wrong sign are excluded.

The results are shown in Fig.~\ref{fig:sterile-hierarchy} for
the three cases 3D CID, 2D CID, and 3D without CID, as earlier.
We may observe 
the following.
\begin{itemize}
  
\item For $|\Delta m^2_{41}| \gtrsim 10^{-2}$ eV$^2$, there is no
  sensitivity to the sterile hierarchy. This is expected, since
  the matter effects are negligible for such large values of $\dm_{41}$ 

\item For $\Delta m^2_{41} \lesssim 10^{-4}$ eV$^2$, there is no
  sensitivity to the sterile hierarchy. This may be attributed to the
  insensitivity of the data to extremely low values of $\dm_{41}$.

\item In the intermediate regime
  $10^{-4}$ eV$^2 \lesssim |\Delta m^2_{41}| \lesssim 10^{-2}$ eV$^2$,
  ICAL is sensitive to the sign of $\dm_{41}$. The sensitivity
  grows as the value of $|\dm_{41}|$ gets closer to $10^{-3}$ eV$^2$.
 This is because at these values of $\dm_{41}$, there is a
  $\nu_\mu$--$\nu_s$ ($\overline{\nu}_\mu$--$\overline{\nu}_s$)
  resonance inside the Earth due to MSW effects,
  for $\dm_{41} <0$ ($\dm_{41}>0$). 
As a result, the neutrino oscillation
  probabilities will be widely different with the two signs of $\dm_{41}$. 

\item  In the same intermediate range,
  three sources contribute to the oscillation probabilities: 
  the oscillation frequencies corresponding to $\dm_{31}$ and $\dm_{41}$,
  as well as the matter potential. The contributions from these sources 
  are of the same order of magnitude in this parameter range.
  The interference of these three contributions results in the smaller
  peaks and valleys as observed in the figure.

\item In addition, it is observed that for $\dm_{41} <0$, the
  $\Delta \chi^2_{\rm SH}$ values are much larger than those for $\dm_{41}>0$.
This may be attributed to the facts that the
MSW resonance occurs in the $\nu_\mu$--$\nu_s$ channel for $\dm_{41}<0$,
and the number of neutrino events at ICAL are almost twice
the number of antineutrino events. 

\item In the absence of hadron energy information (2D CID), the 
 $\Delta \chi^2_{\rm SH}$ values reduce by more than
  25\%, as compared to the 3D CID case, in the intermediate range.
  This illustrates the
  virtue of using the hadron energy information, even though the
  measurement of hadron energy is not very precise at ICAL.

\item Similarly, the absence of CID (3D without CID) would also reduce 
  $\Delta \chi^2_{\rm SH}$ by more than 25\%,
  as compared to the 3D CID case, in the intermediate range.
  This illustrates the advantage ICAL would have due to its
  excellent CID. 

\end{itemize}

The above analysis reinforces the expectation that in the range
$|\dm_{41}| = (0.5-5) \times 10^{-3}$ eV$^2$, the ICAL detector, due to
its capacity to estimate the hadron energy and to determine the muon
charge, would be highly sensitive to the sterile mass hierarchy.

\section{Summary and Concluding Remarks}
\label{sec:conclusions}

The proposed ICAL detector in INO will be able to measure the energy and
direction of muons, induced by the interactions of atmospheric 
$\nu_\mu/\bar{\nu}_\mu$ in the detector, to a very good precision. 
It will also be able to determine the muon charge, hence distinguish the 
incoming muon neutrinos from antineutrinos.   
Its ability to reconstruct the energy of hadron showers on an
event-by-event basis will help it 
observe neutrinos at multi-GeV energies efficiently.
All these features help ICAL to be sensitive to the 
oscillations of atmospheric neutrinos, and to explore
the Earth matter effects on them.

In this paper, we explore the sensitivity of the ICAL detector to the
active-sterile neutrino mixing parameters, in the presence of a single
light sterile neutrino.
If we can confirm the existence of such a light sterile neutrino, 
it would be a revolution in our understanding of neutrinos, as vital
as the discovery of $3\nu$ flavor oscillation.
There is no doubt that exploring the properties 
of such a light sterile neutrino would provide crucial information on the
new physics that is being looked for, at the terrestrial neutrino
oscillation experiments as well as in astrophysics and cosmology.
It is therefore important to explore the sterile neutrino 
parameter space over a wide range of $\dm_{41}$. 
The large $L/E$ range scanned by the atmospheric neutrinos makes ICAL
sensitive to  $\Delta m^2_{41}$ even as low as $10^{-5}$ eV$^2$. 

An important ingredient of this $4\nu$ scenario is the
relative mass squared values, $m_i^2$'s, of the four neutrino mass eigenstates,
i.e. the neutrino mass ordering. While $\dm_{21} >0$, the signs of
$\dm_{31}$ and $\dm_{41}$ lead to four possible mass ordering schemes
(N-N, N-I, I-N, and I-I). Since we would like to explore a wide range of
$\dm_{41}$ values, we have  eight possible mass ordering
configurations (that are the subsets of the above four schemes):
N-N-1, N-N-2, N-N-3, N-I, I-N-1, I-N-2, I-I-1, I-I-2.
One of the main themes in this paper is to quantify the
sensitivity of ICAL to the active-sterile mixing parameters,
for these mass ordering configurations.

Our analysis is based on realistic energy and angular resolutions and
efficiencies for muons induced in CC $\nu_\mu$ interactions,
as obtained by the ICAL collaboration through detector simulations.
The oscillation channels responsible for the signals at ICAL are 
$\nu_\mu \to \nu_\mu$ and $\bar{\nu}_\mu \to \bar{\nu}_\mu$,
and to a smaller extent, 
$\nu_e \to \nu_\mu$ and $\bar{\nu}_e \to \bar{\nu}_\mu$.
The atmospheric neutrino data will therefore be sensitive mainly to 
the mixing parameter $|U_{\mu 4}|^2$ (which affects all the channels
above), with a sub-leading sensitivity to $|U_{e4}|^2$ (which only affects
the appearance channels)\footnote{
There will also be some sensitivity to $|U_{\tau 4}|^2$ due to the
Earth matter effects.}.
We explore the sensitivity of ICAL to the parameters
$\Delta m^2_{41}$ and $|U_{\mu 4}|^2$, while pointing out the changes 
due to the sub-leading contributions from $|U_{e4}|^2$.
In order to achieve this, we solve the neutrino evolution equation
in the four-neutrino framework in the presence of Earth matter,
using the GLoBES package.

In the scenario where no sterile neutrino is present,
we present our sensitivity results in terms of exclusion contours in the
$\Delta m^2_{41}$--$|U_{\mu 4}|^2$ plane, for an exposure of 500 kt-yr.
It is found that the sensitivity reach
of ICAL at very high and very low values of $\Delta m^2_{41}$ is independent of
$\Delta m^2_{41}$. For $\Delta m^2_{41} \gtrsim 1$ eV$^2$, it is due to
the averaging of oscillations; while for 
$\Delta m^2_{41} \lesssim 10^{-4}$ eV$^2$, it is because the oscillations
are governed by $|\dm_{43}| \approx |\dm_{\rm atm}|$. 
In these two ranges, ICAL can exclude the active-sterile neutrino mixing
$|U_{\mu 4}|^2$ up to 0.08 and 0.05, respectively, at 90\% C.L. with  the
two-parameter fit in the 3D analysis mode.
For intermediate $\Delta m^2_{41}$ values, the shapes of exclusion contours
show more interesting features, which are governed by the matter effects
and the interference between atmospheric and sterile mass-squared differences. 
Here the sensitivity is enhanced: for $\Delta m^2_{41} \sim 10^{-3}$ eV$^2$,
values of $|U_{\mu 4}|^2 > 0.02$ can be excluded.
In the range $\dm_{41} \lesssim 0.5 \times 10^{-3}$ eV$^2$, ICAL
can place competitive constraints on $|U_{\mu 4}|^2$ compared to
other existing experiments. 
The exclusion reach is
found to be rather insensitive to the value of $|U_{e4}|^2$ for 
$\Delta m^2_{41} \gtrsim 10^{-3}$ eV$^2$, while at lower $\Delta m^2_{41}$,
nonzero $|U_{e4}|^2$ increases the sensitivity reach of ICAL
for $|U_{\mu 4}|^2$.

We also explore the effects of using only the up(down)-going muon event
samples to identify which kind of events primarily contribute to the
ICAL sensitivity in different $\Delta m^2_{41}$ regions.
As expected, the sensitivity for $\Delta m^2_{41} \gtrsim 1$ eV$^2$ is dominated
by the down-going events, where neutrinos oscillate even over short
distances. 
For low $\Delta m^2_{41} \lesssim 10^{-2}$ eV$^2$, the sensitivity
is dominated by the up-going events. We also find that
at high $\Delta m^2_{41}$, the event rate information is sufficient
to achieve the optimal sensitivity, whereas at low $\Delta m^2_{41}$,
the sensitivity primarily comes from the shape+rate analysis.
Including the hadron energy information in the
$\chi^2$ analysis also helps to improve the sensitivity marginally.
It is observed that the sensitivity reach is  more or less independent
of the mass ordering scheme assumed (N-N, N-I, I-N or I-I).

If a sterile neutrino exists, then the muon charge identification
and hadron energy estimation in ICAL help in the precision measurement
of $\Delta m^2_{41}$, and even the identification of its sign.
We quantify the performance of ICAL in addressing the above
important issues for the first time in this paper.  
For a given true $\dm_{41}=10^{-3}$ eV$^2$, we explicitly calculate
the extent to which test values of $\dm_{41}$ can be excluded.
We find that the mass ordering configurations N-N-1, N-N-3, and N-I
can be ruled out at $\gtrsim 3\sigma$ for the benchmark parameter values.
Since the identification of the mass ordering in the sterile sector
would be one of the priorities of any neutrino physics program
once a sterile neutrino is
discovered, we explore the capability of ICAL to perform this task over a
wide range of possible $\dm_{41}$ values.
We find that for $\dm_{41} \approx (0.5-5.0) \times 10^{-3}$ eV$^2$,
ICAL has a significant capability for identifying the
right mass ordering configuration in the sterile sector.

In summary, ICAL has unique strengths which can play an 
important role in addressing some of the pressing issues 
in active-sterile oscillations, involving a light sterile neutrino 
over a wide mass-squared range. It is one of the few experiments 
sensitive to $\dm_{41} \lesssim 10^{-3}$ eV$^2$ and can put 
strong limits on $|U_{\mu 4}|^2$ in this range.
Its muon charge identification and hadron energy estimation 
capabilities also help in pinning down the magnitude and sign 
of $\dm_{41}$ in the range $(0.5-5.0) \times 10^{-3}$ eV$^2$.
Note that the results obtained in this paper assume muon and 
hadron separation with 100\% efficiency and neglect any 
background. These issues are being addressed by the INO-ICAL 
Collaboration currently. We hope that the analysis performed 
in this paper will take ICAL a step forward in its quest to look 
for new physics beyond the three-neutrino oscillations.

\subsubsection*{Acknowledgments}

This work is a part of the ongoing effort of INO-ICAL collaboration 
to study various physics potentials of the proposed ICAL detector. 
Many members of the collaboration have contributed for the completion 
of this work, especially those who are part of the Physics Analysis meetings.
We are very grateful to K. Bhattacharya, G. Majumder, and 
A. Redij for their developmental work on the ICAL detector simulation package.
We acknowledge the help of J. Kopp for using the new add-on tools
for the GLoBES software, and S.S. Chatterjee for helpful discussions
on the $3\nu$ and $4\nu$ oscillation probabilities.
S.K.A. would like to thank A. Smirnov and A. De Gouvea for useful discussions.
S.K.A. acknowledges the support from DST/INSPIRE Research Grant [IFA-PH-12], 
Department of Science and Technology, India and the Young Scientist Project
[INSA/SP/YSP/144/2017/1578] from the Indian National Science Academy.
M.M.D. acknowledges the support from the Department of Atomic Energy (DAE) 
and the Department of Science and Technology (DST), Government of India,
and the hospitality of the Weizmann Institute of Science, Israel.
A.D. acknowledges partial support from the European Union's Horizon 2020
research and innovation programme under the Marie-Sklodowska-Curie grant
agreement Nos. 674896 and 690575.
T.T. acknowledges support from the Ministerio de Economía y Competitividad
(MINECO): Plan Estatal de Investigación (ref. FPA2015- 65150-C3-1-P,
MINECO/FEDER), Severo Ochoa Centre of Excellence and MultiDark Consolider
(MINECO), and Prometeo program (Generalitat Valenciana), Spain.


\bibliographystyle{JHEP}
\bibliography{ICAL-Sterile-References}

\end{document}